\documentclass[iop,apj]{emulateapj}
\usepackage{amsmath,amssymb,amstext}
\usepackage[version = 4]{mhchem}
\tabletypesize{\scriptsize}

\usepackage[breaklinks,colorlinks,citecolor=cyan,linkcolor=magenta]{hyperref}

\begin{document}

\title{Observations of Disequilibrium CO Chemistry in the Coldest Brown Dwarfs}
\author{Brittany E. Miles}
\affil{University of California, Santa Cruz, Santa Cruz, CA 95064, USA}
\email{bmiles@ucsc.edu}

\author{Andrew J. I. Skemer}
\affil{University of California, Santa Cruz, Santa Cruz, CA 95064, USA}

\author{Caroline V. Morley}
\affil{University of Texas at Austin, Austin, TX 78712, USA}

\author{Mark S. Marley}
\affil{NASA Ames Research Center, Moffett Field, CA 94035, USA}

\author{Jonathan J. Fortney}
\affil{University of California, Santa Cruz, CA 95064, USA}

\author{Katelyn N. Allers}
\affil{Bucknell University, Lewisburg, PA 17837, USA}


\author{Jacqueline K. Faherty}
\affil{American Museum of Natural History, New York, NY 10024, USA}

\author{Thomas R. Geballe}
\affil{Gemini Observatory, Hilo, HI 96720, USA}

\author{Channon Visscher}

\affil{Dordt University, Sioux Center, IA 51250}
\affil{Space Science Institute, Boulder, CO  80301}

\author{Adam C. Schneider}
\affil{Arizona State University, Tempe, AZ 85207, USA}

\author{Roxana Lupu}
\affil{BAER Institute / NASA Ames Research Center, Moffet Field, CA 94035, USA}

\author{Richard S. Freedman}
\affil{SETI Institute, Mountain View, CA 94043, USA}
\affil{NASA Ames Research Center, Moffett Field, CA 94035, USA}

\author{Gordon L. Bjoraker}
\affil{NASA Goddard Space Flight Center, Greenbelt, MD 20771, USA}

\begin{abstract}
Cold brown dwarfs are excellent analogs of widely separated, gas giant exoplanets, and provide insight into the potential atmospheric chemistry and physics we may encounter in objects discovered by future direct imaging surveys. We present a low resolution R $\sim$ 300 $M$-band spectroscopic sequence of seven brown dwarfs with effective temperatures between 750 K and 250 K along with Jupiter. These spectra reveal disequilibrium abundances of carbon monoxide (CO) produced by atmospheric quenching. We use the eddy diffusion coefficient (K$_{zz}$) to estimate the strength of vertical mixing in each object.  The K$_{zz}$ values of cooler gaseous objects are close to their theoretical maximum and warmer objects show weaker mixing, likely due to less efficient convective mixing in primarily radiative layers. The CO-derived K$_{zz}$ values imply that disequilibrium phosphine (\ce{PH3}) should be easily observable in all of the brown dwarfs, but none as yet show any evidence for \ce{PH3} absorption. We find that ammonia is relatively insensitive to atmospheric quenching at these effective temperatures. We are able to improve the fit to WISE 0855's $M$-band spectrum by including both CO and water clouds in the atmospheric model.
\end{abstract}


\section{Introduction}

Studying the methane, ammonia, and water rich atmospheres of late T and early Y- type brown dwarfs is an endeavour that both challenges and improves our understanding of physics and chemistry within the atmospheres of gaseous objects. All current directly imaged planets fall somewhere within the effective temperature range of known brown dwarfs \citep{2016ApJS..225...10F,2016PASP..128j2001B}; the coldest brown dwarfs can be used to forecast the spectroscopic features in colder gas giants we may detect with future direct imaging surveys. We want to know what gas giant planets are made of, how they form, and the processes that take place within their atmospheres. Atmospheric studies of cool brown dwarfs are key to interpreting observations of widely separated gas giant planets.

Y-dwarfs are the coldest type of brown dwarf. They were recently discovered with the \textit{WISE} mission \citep{2011ApJ...743...50C,2012ApJ...753..156K} and several efforts have been made to characterize them using photometry and spectroscopy, primarily across the near infrared. Some of the earliest spectroscopic work by \cite{2011ApJ...743...50C} and \cite{2012ApJ...753..156K} showed that the near infrared absorption features of ammonia (\ce{NH3}), water (\ce{H2O}), and methane (\ce{CH4}) cause the $J$ (1.2 $\mu$m) and $H$ (1.6 $\mu$m) - band peaks to become narrower along the late-T to early-Y dwarf sequence. The main gaseous constituents of late-T and early-Y dwarf atmospheres are known, but atmospheric models often do not fit the available data well \citep{2014ApJ...787...78M, 2014ApJ...783...68B,2015ApJ...804...92S, 2016AJ....152...78L, 2012ApJ...748...74L, 2019arXiv190707798L}. 

The bright thermal background of Earth's atmosphere makes ground-based mid-infrared (3 $\mu$m - 5 $\mu$m) observations of brown dwarfs and gas giant exoplanets challenging, however ``the difficulties of observing in this part of the spectrum are outweighed by the rich variety of molecular bands that are detectable in this interval"\citep{1993ASPC...41...29N}. The 3 $\mu$m - 5 $\mu$m portion of the mid-infrared is where T- and Y- dwarfs emit the majority of their flux and several important gases such as water and methane can be detected \citep{2003ApJ...596..587B,2002Icar..155..393L,2014ApJ...787...78M}. Water absorption was detected in the $M$-band (4.5$\mu$m - 5$\mu$m) spectrum of the coldest brown dwarf, WISE J085510.83-071442.5 (\textbf{WISE 0855}, 250K, \cite{2014ApJ...786L..18L, 2016ApJ...826L..17S}). Methane can be detected in both near infrared and $L$-band (3$\mu$m - 4$\mu$m) spectra of late T-dwarfs (e.g. UGPS 0722, \cite{2012ApJ...748...74L}), but the mid-infrared becomes important for characterizing Y-dwarfs such as WISE 0855 spectroscopically \citep{2018ApJ...858...97M} because these objects emit most of their flux beyond 3$\mu$m (\cite{2014ApJ...783...68B, 2019arXiv190707798L}).Water and methane can be used to estimate the overall atmospheric oxygen and carbon abundances of cool brown dwarfs \citep{2015ApJ...807..183L,2017ApJ...848...83L,2019ApJ...877...24Z} and by extension, gas giant exoplanets that share the same effective temperatures.

Other trace species like carbon monoxide (\ce{CO}), phosphine (\ce{PH3}), arsine (\ce{AsH3}), germane (\ce{GeH4}), and ammonia (\ce{NH3}) could potentially be detected between 3$\mu$m  and  5$\mu$m \citep{2018ApJ...858...97M}, providing constraints on properties such as atmospheric mixing and abundance measurements beyond carbon and oxygen. T- and Y- spectral type brown dwarfs have methane as the dominant carbon-bearing species, but carbon monoxide gas can be brought into the methane-rich regions of the atmosphere through large scale vertical mixing \citep{2002Icar..155..393L}. Disequilibrium carbon monoxide abundances have been inferred photometrically and confirmed spectroscopically in several T dwarfs providing evidence of atmospheric quenching driven by convective mixing \citep{1997ApJ...489L..87N, 1998ApJ...502..932O,2003IAUS..211..345S, 2004AJ....127.3516G,2009ApJ...695..844G, 2012ApJ...760..151S, 2012ApJ...748...74L}. The coldest T-dwarf with a disequilibrium \ce{CO} detection is Gl 570 D, which has a near infrared classification as a T8 \citep{2006ApJ...637.1067B}.

Phosphine is another signal of strong convective mixing that has been observed in Jupiter (126 K), but it was not seen in WISE 0855's (250 K) $L$ or $M$-band spectrum \citep{2016ApJ...826L..17S, 2018ApJ...858...97M}. Jupiter and WISE 0855 are nearly similar in temperature (126 K vs 250 K), but phosphine's abundance is only quenched within Jupiter's atmosphere. One of the goals of this work is to understand the diversity of atmospheric mixing between the parameter space of late-T dwarfs to early Y-dwarfs to giant planets.

Water clouds are predicted to be significant sources of opacity in brown dwarfs with effective temperatures below $\sim$375 K \citep{2003ApJ...596..587B, 2014ApJ...787...78M}. WISE 0855's $M$-band spectrum shows evidence of water clouds, because the spectral shape cannot be fit with a cloudless model \citep{2016ApJ...826L..17S,2018ApJ...869...18M}. Cloudy models work well for fitting WISE 0855's normalized $M$-band spectra in isolation, but matching the available photometry is still an issue that could potentially be resolved with the consideration of upper atmospheric heating or other opacity sources  \citep{2016ApJ...832...58E,2018ApJ...858...97M, 2019ApJ...882..117L}. 

Extending the previous work of \cite{2016ApJ...826L..17S} and \cite{2018ApJ...858...97M}, we explore the atmospheric properties of seven brown dwarfs with effective temperatures covering 700 K to 250 K and Jupiter based on $M$-band spectroscopic data. In this study we present new Gemini/GNIRS $M$-band spectral observations of WISEPA J031325.96+780744.2 (\textbf{WISE 0313}, \cite{2011ApJS..197...19K}), UGPS J072227.51-054031.2 (\textbf{UGPS 0722}, \cite{2010MNRAS.408L..56L}), WISEPC J205628.90+145953.3 (\textbf{WISE 2056}, \cite{2011ApJ...743...50C}), and WISEP J154151.65-225025.2 (\textbf{WISE 1541}, \cite{2011ApJ...743...50C}). These new data are supplemented by previously published \textit{AKARI} observations of \textbf{2MASS J0415-0935} and Gemini/NIRI observations of  \textbf{Gl 570 D}  \citep{2012ApJ...760..151S, 2009ApJ...695..844G}. The WISE 0855 $M$-band data presented in \cite{2016ApJ...826L..17S} are re-reduced using the methods in this work and added to the sample. All spectra are fit better with disequilibrium \ce{CO} abundances, even though methane is the dominant carbon bearing gas at these effective temperatures.

These observations address two questions: \textbf{1)} What do cool brown dwarfs look like spectroscopically across the $M$-band and why? \textbf{2)} Are there any trends in atmospheric quenching within this effective temperature range?

The observations and data reduction methods for the new $M$-band data are described in Sections~\ref{sec:observations} and~\ref{sec:reduction}. In Section~\ref{sec:sequence}, the $M$-band spectral sequence ordered by luminosity derived effective temperatures is shown. We show that equilibrium models do not adequately recreate the features of the spectra and disequilibrium abundances of carbon monoxide are needed to fit the spectra in Section~\ref{sec:model_comparison}. We briefly discuss the effect of carbon monoxide and clouds in WISE 0855 and WISE 1541 and their improvement of the spectral fits. In Section~\ref{sec:atm_mixing} we estimate the eddy diffusion coefficient for each object based on carbon monoxide abundances and make predictions for phosphine and ammonia, which are summarized in Table~\ref{tbl:quench abundances}. Lastly we discuss the implications of this work for cold, directly imaged gas giant exoplanets that may be discovered in the future. 

\section{Observations}
\label{sec:observations}

We obtained observations of WISE 0313, UGPS J0722, WISE 2056, and WISE 1541 over the course of a year at Gemini North (Programs GN-2016B-Q-23, GN-2017A-Q-5, GN-2017A-Q-32)  using the Gemini near-infrared spectrograph (GNIRS; \citealt{2006SPIE.6269E..4CE}). These observations are complementary to the $M$-band spectra of WISE 0855 published in \cite{2016ApJ...826L..17S}.

GNIRS was set up with the long camera (0.05$\arcsec$/pix  resolution), 0.675$\arcsec$ slit, and deep detector well depth setting for all observations. There is uncertainty in the position of our objects due to their relatively high proper motions. We use a 13.5 pixel wide (0.675$\arcsec$) slit to avoid missing the objects completely when placing them in the slit. The deep well setting is needed to record more of the bright sky background without reaching non-linearity within the detector pixels. The 31.7 line mm$^{-1}$ grating covers the $M$-band from 4.5 $\mu$m to 5.1 $\mu$m at an effective resolution of $\sim$ 7,400, which is eventually binned down to a resolution of $\sim$370. Each spectral image is the sum of 24 co-added 2.5 second long integrations. The total integration time was 4.2 hours for WISE 0313, 2.4 hours for UGPS 0722, 15.36 for WISE 2056, and 10.8 hours for WISE 1541.

A single observation sequence of a target consists of four spectral images in a ABBA pattern by nodding along the slit. Each block of data typically takes the following pattern:

\begin{enumerate}
    \item Acquisition of first telluric calibrator star
    \item First telluric calibrator star observation sequence. 
     \begin{sloppypar}
        (1 x 24 co-adds X 2.5 seconds)
     \end{sloppypar}
    \item Acquisition of brown dwarf
    \item Nine brown dwarf observation sequences.
    \begin{sloppypar}
        (9 x 24 co-adds X 2.5 seconds)
     \end{sloppypar}
    \item Re-acquisition of brown dwarf.
    \item Nine brown dwarf observation sequences.
     \begin{sloppypar}
        (9 x 24 co-adds X 2.5 seconds)
     \end{sloppypar}
    \item Acquisition of second telluric calibrator star
    \item Second telluric calibrator star observation sequence.
    \begin{sloppypar}
        (1 x 24 co-adds X 2.5 seconds)
     \end{sloppypar}
\end{enumerate}

Every brown dwarf target was acquired by blind-offsetting from a bright, nearby star except for UGPS 0722. The position of the brown dwarf was calculated into a standard reference frame using proper motion propagation code and published parallaxes (UGPS 0722 - \citealt{2012ApJ...753..156K}, WISE 0313 - \citealt{2014ApJ...783...68B}, WISE 2056 - \citealt{2019ApJS..240...19K}, WISE 1541 - \citealt{2019ApJS..240...19K}). Every telluric calibrator is an A0 or A1 type star that is nearby and within .2 airmasses of its respective science target on the sky. The before and after calibrator stars are kept the same for an individual brown dwarf in all programs. A full breakdown of observations used for each object are listed in Table~\ref{tbl:obs-gemini}.

\clearpage
\begin{turnpage}
\begin{deluxetable*}{cccccclcccc} 
\caption{}
\tablecaption{Summary of GNIRS Observations.}
\tabletypesize{\tiny} 
\tablehead{
\colhead{Science} &  \colhead{Date} &   \colhead{Total Integration}   & \colhead{Science}  & \colhead{Science} & \colhead{Science} & \colhead{Telluric}   & \colhead{Telluric} & \colhead{Telluric} & \colhead{Telluric} & \colhead{Data Comments} \\
\colhead{Target}  &  \colhead{YYYYMMDD} &  \colhead{ Time (minutes) } &  \colhead{Airmass} & \colhead{IQ}      & \colhead{WVC}     & \colhead{Calibrator} &  \colhead{Airmass} & \colhead{IQ}        & \colhead{WVC} & \colhead{}
}
\startdata
WISE 0313 & 20161014 & 36.0 & 1.899  & 70\% & 20\%  & HIP 10054 & 2.090 & 70\% & 20\% &  \\
WISE 0313 & 20161014 & 36.0 & 1.934  & 70\% & 20\% & HIP 16725 & 1.746 & 70\% & 20\% & The single science frame with detector noise not included. \\
WISE 0313 & 20161015 & 36.0 & 1.909 & 70\% & 50\%  & HIP 10054 & 2.088 & 70\% & 50\% & \\
WISE 0313 & 20161015 & 36.0 & 1.901  & 70\% & 50\% & HIP 16725 & 1.691 & 70\% & 50\% & \\
WISE 0313 & 20161104 & 36.0 & 1.973 & 70\% & 50/80\%  & HIP 10054 & 2.109 & 70\% & 50\% & \\
WISE 0313 & 20161104 & 24.0 & 2.028 & 70\% & 50\%  & HIP 16725 & 1.827 & 20\% &  50\% & \\
WISE 0313 & 20170103 & 12.0 & 1.915  & 70\%  & 50\%  & HIP 10054 & 2.093  & 70\% & 50\% & \\
WISE 0313 & 20170103 & 12.0 & 1.899  & 20\%  & 50\%  & HIP 16725 & 1.687 & 20\% & 50\% & \\
\hline
UGPS 0722 & 20161115 & 36.0 & 1.110 & 70\%  & 20\% &  HIP 30387 & 1.104 & 70\%  & 20\% &  \\
UGPS 0722 & 20161115 & 36.0 & 1.164 & 70\%  & 20\%  &  HIP 42028 & 1.093 & 70\%  & 20\% & \\
UGPS 0722 & 20170104 & 36.0 & 1.186 & 70\% & 50\% &  HIP 30387 & 1.152 & 20\%  & 50\% & \\
UGPS 0722 & 20170104 & 36.0 & 1.110 & 20\%  & 50\% &  HIP 42028 & 1.101 & 70\% & 50\% & \\
\hline
WISE 2056 & 20161009 & 36.0 & 1.057  & 20\%  & 80\%  & HIP 95002 & 1.173 & 20\% & 80\% &  \\ 
WISE 2056 & 20161009 & 36.0 & 1.207 &  20\% & 80\% & HIP 108060 & 1.135 & 20\% &  80\% &\\
WISE 2056 & 20161010 & 36.0 & 1.014 &  20\%  & 80\% & HIP 95002 & 1.013 & 20\% & 80\%  & 2056 traces not visible. Brown dwarf data not included\\
WISE 2056 & 20161010 & 36.0 & 1.011 & 20\%  &  80\% & HIP 108060 & 1.002 & 20\% & 80\% & Detector noise. Brown dwarf data not included \\
WISE 2056 & 20161011 & 36.0 & 1.034  & 70\%  & 50\%  & HIP 95002 & 1.113 & 70\% &50\%  &\\
WISE 2056 & 20161011 & 36.0 & 1.231 & 70\%  & 50\%  & HIP 108060 & 1.156 & 70\% & 50\% & 2056 traces not visible, Detector noise. Brown dwarf data not included.\\
WISE 2056 & 20161014 & 36.0 & 1.006 & 70\%  & 50\%  & HIP 95002 & 1.023 & 70\% & 50\% & 2056 traces not visible, Brown dwarf data not included.\\
WISE 2056 & 20161014 & 36.0 & 1.026 & 70\% &  50\% & HIP 108060 & 1.008 & 70\% & 50\% & \\
WISE 2056 & 20161015 & 36.0 & 1.006 &70\%  & 50\% & HIP 95002 &  1.026 & 70\% & 50\% &\\
WISE 2056 & 20161015 & 36.0 & 1.021 & 70\%  & 50\%  & HIP 108060 & 1.006 & 70\% & 50\% & \\
WISE 2056 & 20161018 & 36.0 & 1.005  & 70\%  &  20\% & HIP 95002 &  1.026 & 70\% & 20\%/50\% & Fog, detector noise. Brown dwarf data not included\\
WISE 2056 & 20161018 & 22.0 & 1.015  & 70\%  & 20\%  & HIP 95002 & 1.026 & 70\% & 20\%/50\%  & Fog, detector noise. Brown dwarf data not included\\
WISE 2056 & 20161022 & 36.0 & 1.005 & 20\%  & 80\%  & HIP 95002 & 1.045 & 20\% & 80\% &\\
WISE 2056 & 20161022 & 36.0 & 1.042  & 70\%  & 80\%  & HIP 108060 & 1.020 & 70\% & 80\% &\\
WISE 2056 & 20161113 & 36.0 & 1.064  & 70\% & 50\%  & HIP 95002 & 1.206 & 70\% & 50\% & Calibrations taken in K-band. Brown dwarf data not included\\
WISE 2056 & 20161113 & 36.0 & 1.215  & 70\%  & 50\%  & HIP 108060 & 1.162 & 70\% & 50\% & Detector noise. Brown dwarf data not included\\
WISE 2056 & 20161114 & 36.0 & 1.058  & 70\%  & 50\%  & HIP 95002 & 1.1985 & 70\% &50\% &\\
WISE 2056 & 20161114 & 36.0 & 1.193  & 70\% & 50\%  & HIP 108060 & 1.127 & 70\% & 50\% &\\
WISE 2056 & 20161115 & 36.0 & 1.070  & 70\%  & 50\%  & HIP 95002 & 1.222 & 70\% & 80\% &\\
WISE 2056 & 20161115 & 36.0 & 1.224 & 70\%  & 50\%  & HIP 108060 & 1.158 & 70\% & 20\% & \\ 
WISE 2056 & 20170611 & 36.0 & 1.099 & 20\% & 80\%  & HIP 95002 & 1.030  & 70\% & 80\% &\\
WISE 2056 & 20170611 & 36.0 & 1.018 & 20\%  & 80\% & HIP 108060 & 1.024 & 70\% & 80\% &\\
WISE 2056 & 20170612 & 36.0 & 1.267  & 20\%  & 80\%  & HIP 95002 & 1.112 & 20\% &80\%  & \\
WISE 2056 & 20170612 & 36.0 & 1.087  & 20\%  & 80\%  & HIP 108060 & 1.113  & 70\% & 80\% &\\
WISE 2056 & 20170616 & 36.0 & 1.221 & 70\%  & 80\%  & HIP 95002 & 1.087 & 70\% & 80\% &\\
WISE 2056 & 20170616 & 36.0 & 1.067  & 70\% & 80\%  & HIP 108060 & 1.086 & 70\% & 80\% &\\
\hline
WISE 1541 & 20170517  & 36.0 & 1.363  & 70\%  & 20\% & HIP 70765 & 1.335 & 70\% &  20\% &\\
WISE 1541 & 20170517  & 36.0 & 1.422 & 70\% & 20\%  & HIP 81457 &  1.437 & 70\% & 20\% & Severe detector noise. Brown dwarf data not included\\
WISE 1541 & 20170518  & 36.0 & 1.363  & 70\%  & 20\%  & HIP 70765 & 1.324 & 70\% & 20\% &\\
WISE 1541 & 20170518  & 36.0 & 1.436  & 70\% & 20\%  & HIP 81457 & 1.455 & 70\% & 20\%  &\\
WISE 1541 & 20170605  & 36.0 & 1.394  & 70\%  & 80\% & HIP 70765 & 1.316 & 70\% & 80\%  &\\
WISE 1541 & 20170605  & 36.0 & 1.363  & 70\%  & 80\%  & HIP 81457 &  1.406 & 70\% & 80\% &\\
WISE 1541 & 20170606  & 36.0 & 1.368 & 70\%  & 50\%  & HIP 70765 & 1.315 & 70\% & 50\%  &\\
WISE 1541 & 20170606  & 36.0 & 1.378 & 70\% & 50\%  & HIP 81457 &  1.414 & 70\% & 50\%  &\\
WISE 1541 & 20170610  & 36.0 & 1.388  & 70\%  & $>$80\%  & HIP 70765 & 1.316 & 70\% &  80\%  &\\
WISE 1541 & 20170610  & 36.0 & 1.365  & 20\%  & $>$80\%  & HIP 81457 & 1.407 & 20\% &  80\% &\\
WISE 1541 & 20170611  & 36.0 & 1.428  &  70\% & 50\% / 80\% & HIP 70765 & 1.327 & 70\% & 80\%  &\\
WISE 1541 & 20170611  & 36.0 & 1.362 & 70\%  & 80\%  & HIP 81457 & 1.408 & 70\% & 80\% & \\
WISE 1541 & 20170612  & 36.0 & 1.414  &  20\%  & 80\%  & HIP 70765 & 1.328 & 20\% & 80\% & WISE 1541 traces not visible. Brown dwarf data not included \\
WISE 1541 & 20170612  & 36.0 & 1.362  & 20\% & 80\%  & HIP 81457 & 1.410 & 20\% &80\%  &\\
WISE 1541 & 20170623  & 36.0 & 1.451  & 20\%  &  80\% & HIP 70765 & 1.336 & 20\% & 80\%  &\\
WISE 1541 & 20170623  & 36.0 &  1.363  & 20\%  & 80\%  & HIP 81457 & 1.426 & 20\% & 80\%  &\\
WISE 1541 & 20170708  & 36.0 & 1.362  &  20\%  & 20\%  & HIP 70765 & 1.329 & 20\% & 20\%  &\\
WISE 1541 & 20170708  & 36.0 & 1.416  & 20\%  & 20\% & HIP 81457 & 1.432 & 70\% & 20\%  & 
\tablecomments{The reported air masses of the science target and calibrator observations are median values over the course of the sequence. Image Quality (IQ) and Water Vapor Content (WVC) describe constraints on the point spread function width and the water content of the observations. See http://www.gemini.edu/sciops/telescopes-and-sites/observing-condition-constraints for more details}
\label{tbl:obs-gemini}
\end{deluxetable*}
\end{turnpage}

\section{Data Reduction}
\label{sec:reduction}

A single subset of spectral data consists of a telluric calibrator sequence and nine science target sequences. A subset is reduced to create two extracted science spectra. The final spectrum for each science target is the combination of all the extracted spectra from each subset of that object.

\subsection{Spectral Images Removed from Analysis}
On occasion, the Gemini NOAO Aladdin Array Controller\footnote{http://www.gemini.edu/sciops/instruments/gnirs/known-issues} will create a checkerboard pattern visible in the A nod minus B nod (A-B) frames of calibrator and science targets. Spectral images affected by this noise are removed from the analysis. If the majority of a brown dwarf's observation sequence is affected by electronic pattern noise, it is not included in the analysis. Observations affected by electronic pattern noise are noted in the comments of Table~\ref{tbl:obs-gemini}.

Despite the short integration times used for our observations, some longer-wavelength skylines still reach non-linearity. Wavelengths with count values in the non-linear range are flagged and later removed during the spectral extraction process (See Section \ref{subsec:non-lin}). Clouds create a bright and variable sky background that eclipses the science target signal, making A-B sky subtraction useless. The 2016 - 10 - 18 observations of WISE 2056 are excluded from the analysis due to this.

\subsection{Spectral Image Reduction}

Sections 3.2 and 3.3 of \cite{2018ApJ...869...18M} based on the REDSPEC package \citep{2015ascl.soft07017K} outline the procedure used for the sky subtraction, rectification, and wavelength solution of the spectral image data. These procedures are similar to the ones used to reduce the WISE 0855 data presented in \cite{2016ApJ...826L..17S}. Centroids  are  fit  along  the  traces of the mean A-B telluric calibrator spectral images to estimate the deviation from a straight line and create the spatial rectification map used for interpolation. Rectification maps have coefficients for each row used for one-dimensional interpolation.\footnotetext{numpy.interp} The spatial rectification map is applied to a mean A nod plus B nod (A+B) image of the telluric calibrator, where 12 sky lines are used to create a second rectification map for the wavelength direction. The A-B spectral images of the telluric calibrator and science target are taken to subtract most of the sky background. Each of these differenced spectral images are then interpolated using the rectification maps derived from the mean calibrator image. Excess sky remains after nod subtracting and rectifying the spectra, therefore at each row along the entire wavelength direction, the median of the pixels along the spatial direction is subtracted.
 
Following the ordering of the observational pattern steps listed in Section~\ref{sec:observations}, the rectification and wavelength maps from the before telluric calibrator (Step 2) are applied to the first brown dwarf's sequence (Step 4). The rectification and wavelength maps from the after telluric calibrator (Step 8) are applied to the second brown dwarf sequence (Step 6). 

The wavelength solution for each subset is calculated by fitting a second order polynomial to 13 sky emission lines along the wavelength direction of the spatially and spectrally rectified A+B calibrator images. The spectral features are identified by referencing a model of the Maunakea Sky\footnote{www.gemini.edu/sciops/telescopes-and-sites/ \\ observing-condition-constraints/ir-background-spectra} smoothed to the resolution of the data.

\subsection{Detector Non-Linearity}
\label{subsec:non-lin}

Earth's atmosphere has strong background emission across the $M$-band which can drive the detector into a non-linear regime. A very conservative approach is taken to address non-linearity on the detector for a better telluric calibration. Pixels with values in the non-linear response regime of $>$10,000 ADU per co-add\footnote{https://www.gemini.edu/sciops/instruments/gnirs/spectroscopy/detector-properties-and-read-modes} are marked in every spectral image by creating a separate non-linearity image where 0 is linear and 1 is non-linearity. These non-linearity images are also rectified using the appropriate rectification map. Interpolation causes surrounding pixels to be flagged (values above 0) in the non linear regime even if they were not originally. At most, 3.4\% more pixels are flagged in the rectified non-linear maps than in the original spectral image. For a single subset (calibrator and science), the mean of all of the non-linearity maps are taken to create a final single map for the entire subset. If any pixel was flagged as non-linear over the course of a subset it is masked from the analysis.

\subsection{Spectral Extraction and Error Estimation}

After sky subtraction and rectification, a 3$\sigma$ clip is made for each pixel along the stack of reduced science images in a subset to remove outliers. The mean of the reduced science image is used for extraction. The reduced mean science image is collapsed over the spatial direction by taking a mean weighted by noise in order to find the positive and negative traces of the faint science target. The boxcar extraction center and radius (1.5852 $\sigma$)\footnote{http://wise2.ipac.caltech.edu/staff/fmasci/GaussApRadius.pdf} of each trace are estimated by using the best-fit parameters of a Gaussian. The same extraction procedure is applied to the reduced calibrator images.

The errors are estimated by taking the variance image of a reduced stack of images and doing a boxcar extraction over the same center and widths as the respective trace. The rectified non-linear maps created for each subset (calibrator and science target pointing) are also boxcar extracted to find the wavelengths affected by non-linear pixels. If any pixel at a given wavelength within an extraction width had a non-linear response, the entire wavelength element is masked out in the final extracted spectrum. 

\subsection{Telluric and Relative Flux Calibration}
Calibrator spectra of A0 and A1 stars are taken to remove the response of the telescope and Earth's atmosphere because the stellar spectra can be reproduced by a black body function across the mid-infrared. The extracted science spectra are divided by a calibrator star and then multiplied by Planck's law using the same temperature of the calibrator star. The calibrator star HIP 39898 from the WISE 0855 program \citep{2016ApJ...826L..17S}  has a visible Pfund hydrogen recombination line at 4.65 $\mu$m. This line is removed prior to division by fitting a Gaussian to the feature. No calibrator stars from programs GN-2016B-Q-23, GN-2017A-Q-5, GN-2017A-Q-32 show Pfund (7 $->$ 5, 4.65 $\mu$m) or Humphreys (11 $->$ 6, 5.12 $\mu$m) hydrogen emission across the $M$-band.

Spectra of A0 stars are taken at the beginning and end of each observational block to assess the quality of the telluric calibration. We extract the calibrator spectra taken before and after each science target observational block, normalize by the median, and then divide the before calibrator by the after calibrator. The ratio of the normalized ``before" calibrator to the normalized ``after" calibrator should be one, however there are deviations which are interpreted as percentage errors included in the final spectrum.

\begin{figure*}
\centering
\includegraphics[width=7in]{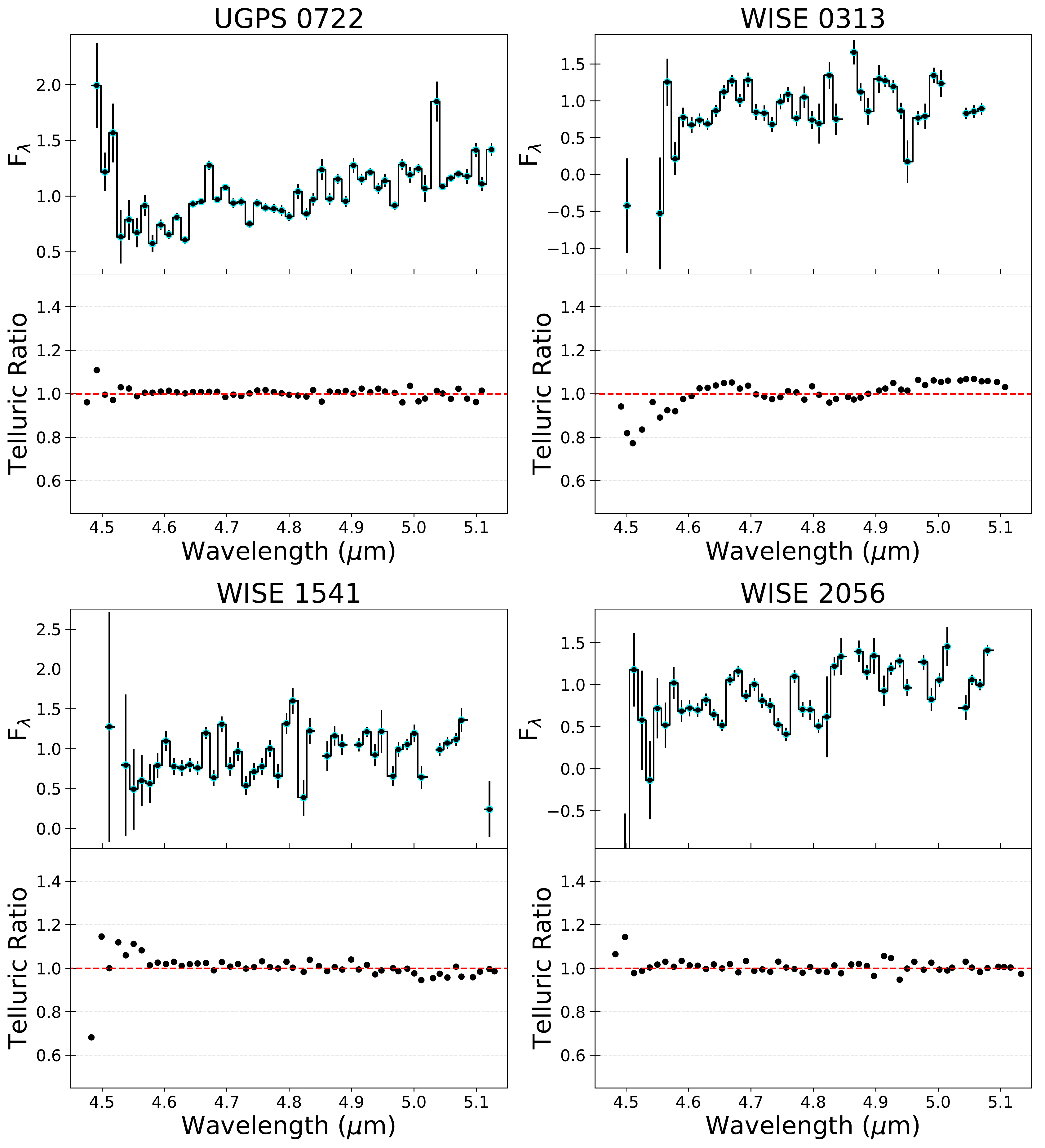}
\caption{Each brown dwarf has a panel with a spectrum ($Top$) and the combined telluric ratio over the course of the program ($Bottom$). The telluric ratio is a metric for how consistent a calibrator spectrum is over a 2 hour period when the science data are taken. Percent deviations from unity are folded into the error bars of the final science spectrum of each object. The ratio between the before and after calibrators departs from unity significantly near areas of low atmospheric transmission. The median deviation for all brown dwarf telluric ratios is less than 2\%, the largest deviations occur shortward of 4.55 $\mu$m where the deviations are as high as 14.5\%}
\label{fig:m-band-spectra}
\end{figure*}

To get the final spectrum for each object, the normalized spectra from every subset are placed into a single array, then the data points are re-ordered by wavelength.  The data points are then binned using an average weighted by the error to produce a 51 pixel length spectrum with a resolution of about 370 of each brown dwarf. $M$-band observations are background limited and binning always increases the signal-to-noise per pixel along the spectrum. The original WISE 0855 $M$-band spectrum was binned to 51 wavelength elements to flatten out ringing variations in the telluric ratio. The objects in this work are binned to the same amount of wavelength elements for convenience. The spectra of UGPS 0722, WISE 0313, WISE 1541, and WISE 2056 are shown in Figure~\ref{fig:m-band-spectra}.

\subsection{Reduction Comparison with Skemer et al. 2016 }

\begin{figure}
\centering
\includegraphics[width=3in]{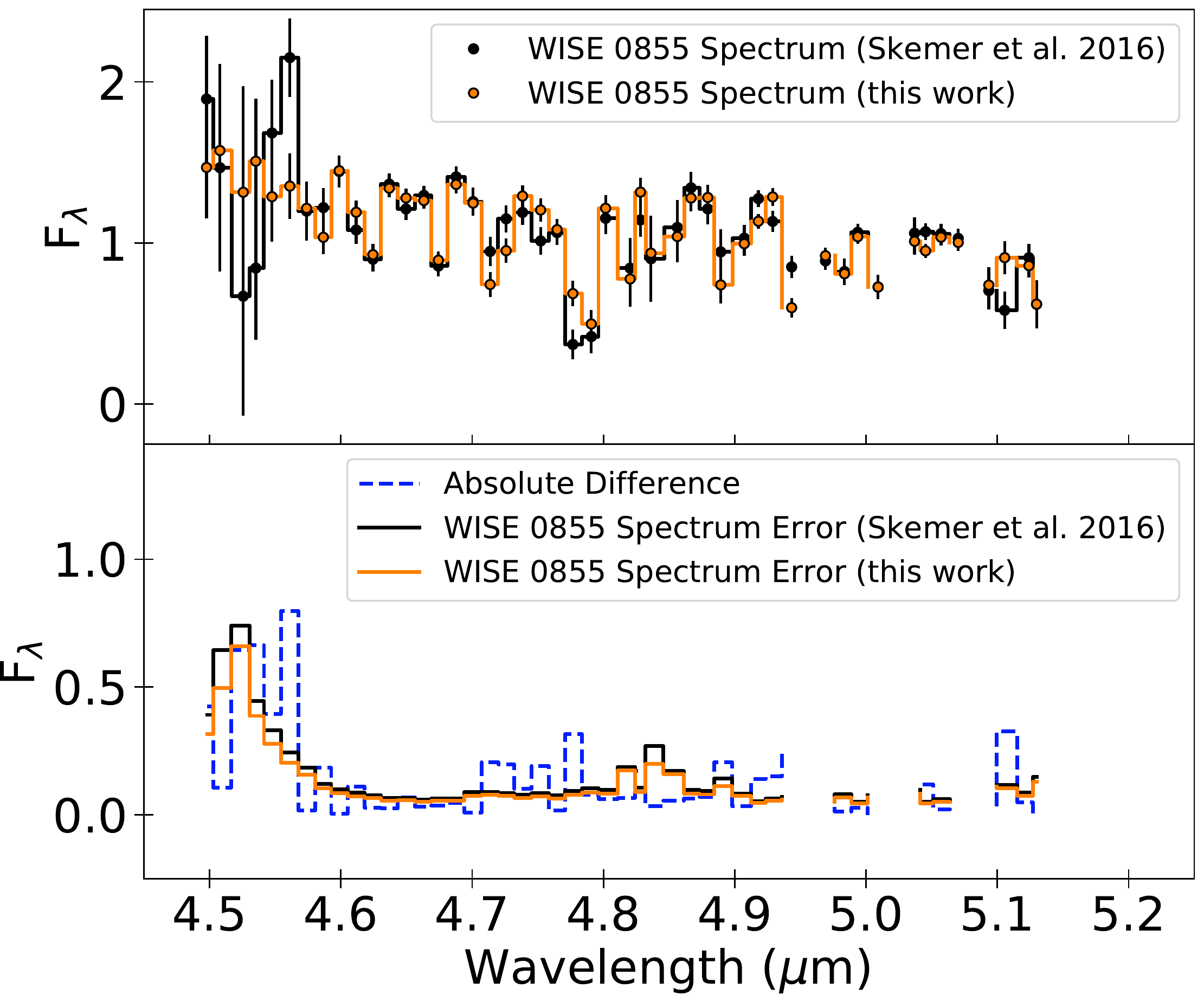}
\caption{$Top$ $Panel$ -  Black - The final WISE 0855 spectrum presented in \cite{2016ApJ...826L..17S}, Orange - The same data from \cite{2016ApJ...826L..17S} re-reduced using the methods described in this paper and interpolated onto the same wavelength spacing as the \cite{2016ApJ...826L..17S} spectrum. Error bars are plotted in black. $Bottom$ $Panel$ - Blue Dashed Line - The absolute difference between the previously published spectrum and the re-reduced spectrum. The solid orange and black lines are the errors of each spectrum. The spectral shapes are consistent, however there are discrepancies especially within regions of relatively low signal to noise.}
\label{fig:comparision}
\end{figure}

The WISE 0855 data published in \cite{2016ApJ...826L..17S} were re-reduced using the procedure described in this work. The results are compared in Figure~\ref{fig:comparision}. In \cite{2016ApJ...826L..17S}, the pixels affected by non-linearity 10\% or less of the time in a mean spectral image were used in the extraction process. In this work, no non-linearity is accepted and this primarily affects the redder portion of the $M$-band spectrum where the sky is relatively brighter. On the blue side of the $M$-band spectrum there are fewer sky lines, leading to interpolation differences between this work and \cite{2016ApJ...826L..17S}. Qualitatively, the two reductions show very similar spectral shapes and absorption features across the $M$-band. The spectral points are on average within 1$\sigma$ of each other, the largest differences are often in low signal-to-noise areas but the discrepancies are less than 3.5$\sigma$.

\section{Sequence of Cool Objects}
\label{sec:sequence}

\subsection{Object Temperatures}
Brown dwarfs are often classified by their near-infrared spectral types rather than effective temperatures inferred from models, because the molecular features associated with certain spectral types can be proxies for effective temperature \citep{2005ARA&A..43..195K}. Near-infrared spectral indicators have been used to infer properties such as surface gravity, but these indicators are challenging to measure against the faint near-infrared emission of T and early Y dwarfs \citep{2013ApJ...772...79A, 2017ApJ...838...73M}. Model-derived properties do have drawbacks, but they can be useful for predicting the gases or condensates that should be observable and constraining other atmospheric characteristics like vertical mixing, weather, and climate.  Each object is classified by the effective temperature derived from estimated total fluxes and evolution models. 

The adopted effective temperatures are derived from published near- and mid-IR photometry which captures 50 \% to 84 \% of the emitted flux from these brown dwarfs. Published near-infrared spectra could be used to constrain effective temperatures, but they do not cover the wavelength range where cooler brown dwarfs emit most of their luminosity and the spectra tend to dominate the fitting process. The Sonora Bobcat grid evolution models (Marley et al. in prep.)\footnotetext{Sonora Bobcat grid models - https://zenodo.org/record/2628068\#.Xb99ESV7lTI} of solar metallicity, cloudless models are used to find a range of possible effective temperatures for each brown dwarf. The evolution model grid covers ages from 6000 years to 1 Gyr and masses of 0.5 to 102 Jupiter masses. We limit the mass range of the models to be between 1 Jupiter mass and 83 Jupiter masses.  The 83 Jupiter mass upper limit is an estimate on the boundary between hydrogen-burning stars and brown dwarfs. Model ages are limited to cover from 1 Gyr to 10 Gyr; the only exception to this is UGPS 0722, whose inferred age range is 60 Myr and 1 Gyr \citep{2012ApJ...748...74L} based on kinematics. Gl 570D is assumed to be over 1 Gyr due the lack of activity from a stellar companion in the system \citep{2000ApJ...531L..57B}. Gl 570 D and 2M0415 are assumed to have age upper limits of about 10 Gyr due their kinematics placing them in the Milky Way's galactic disk\citep{2000ApJ...531L..57B,2007ApJ...656.1136S}. WISE 0313, WISE 2056, WISE 1541, and WISE 0855 do not have age constraints, but are assumed to have ages between 1 and 10 Gyr because they are all within 10 parsecs of the Sun like Gl570 D and 2M0415. The effective temperature range is determined by the evolution models that equal the estimated total flux derived from photometry. The adopted effective temperature is the Sonora grid atmospheric model with the effective temperature closest to the mean of the physical range. The Sonora Bobcat grid atmospheric model covers effective temperatures between 200 K  - 2200 K with 25 K increments below 600 K, 50 K increments below 1000 K and 100 K increments below 2000 K. The surface gravities span log(g) = 3.2 to log(g) = 5.5 with increments of .25. Jupiter's effective temperature of 126 K is adopted from \cite{2012JGRE..11711002L}. 

\begin{figure*}
\centering
\includegraphics[width=6in]{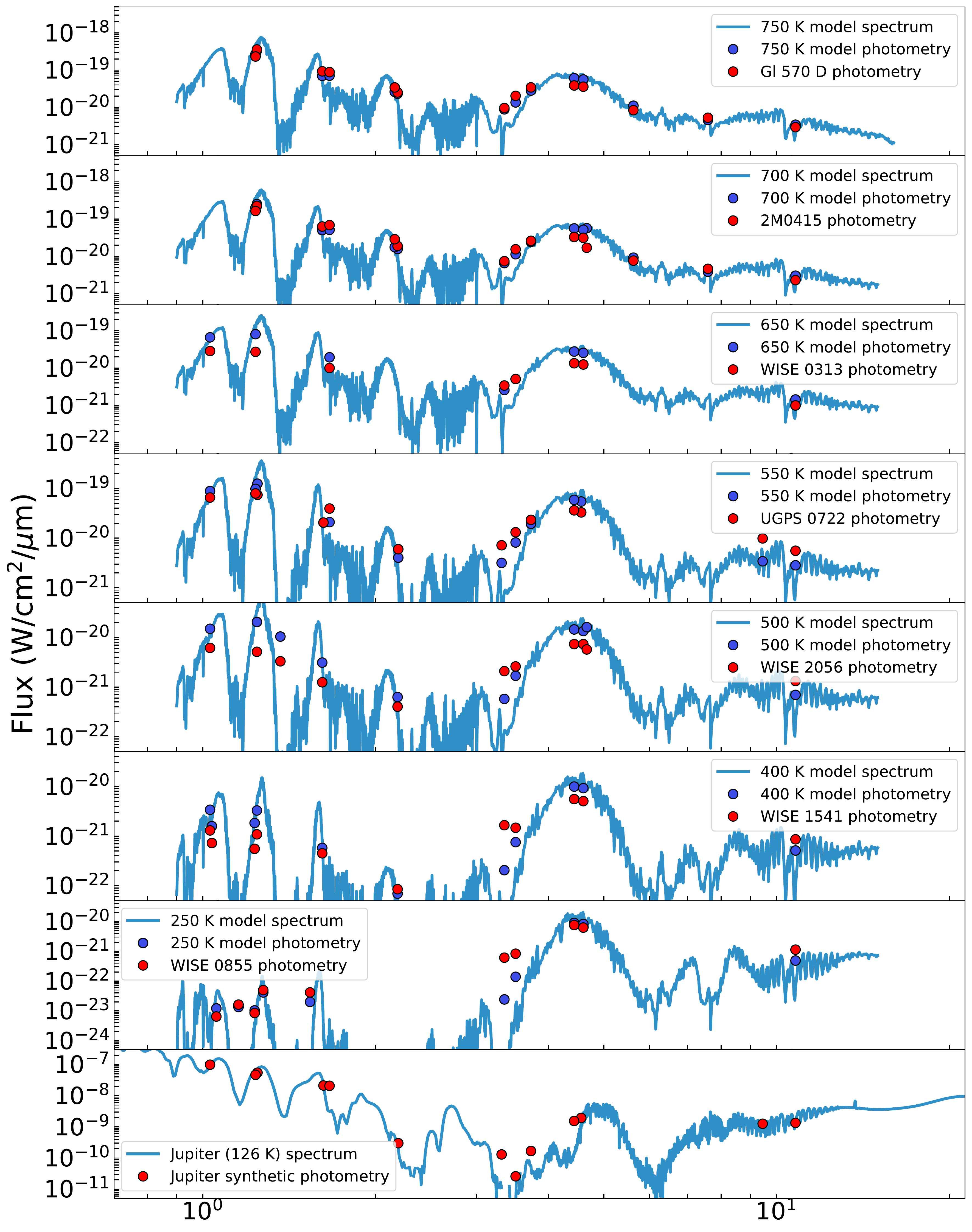}
\caption{Model spectra (Blue lines) of the adopted temperatures scaled to the appropriate distance of the respective brown dwarf assuming thermochemical equilibrium, a size of 1 Jupiter radius, log(g) = 4.5, no clouds, and solar metallicity. The red points are the measured photometry of the brown dwarfs, blue points are the photometry derived from the model spectra. The error bars are about the size of the photometry points. There are major discrepancies across the near and mid-infrared, but the majority (50\% - 89\%) of the luminosity is captured with the available photometry. Jupiter's spectrum is real data was compiled by Mike Cushing (private communication) and originally taken with the Cassini Composite Infrared Spectrometer (CIRS), Infrared Space Observatory (ISO) Short Wavelength Spectrometer (SWS), and the Galileo Near-Infrared Mapping Spectrometer (NIMS). Shortward of 4 microns, Jupiter's spectrum is entirely reflected sunlight and without the Sun, the \ce{CH4}, \ce{C2H6}, and \ce{C2H2} emission lines would not be present.  The synthetic photometry of Jupiter is calcuated using the same bandpasses as the UGPS 0722 data.}
\label{fig:photometry}
\end{figure*}

The list of substellar objects studied along with their published and adopted properties are shown in Table~\ref{tbl:objects}. The adopted effective temperature model spectra and synthetic photometry are compared against the published photometry in Figure~\ref{fig:photometry}. All of the estimated effective temperature ranges are consistent with previously published temperatures (Gl 570D \citealt{2000ApJ...531L..57B,2004AJ....127.3516G,2015ApJ...810..158F}, 2M0415 -  \citealt{2004AJ....127.3516G,2007ApJ...656.1136S,2015ApJ...810..158F}, WISE 0313 - \cite{2014ApJ...783...68B}, UGPS 0722 - \citealt{2010MNRAS.408L..56L,2012ApJ...748...74L,2015ApJ...810..158F}, WISE 2056 -  \citealt{2014ApJ...783...68B,2017ApJ...842..118L,2019ApJ...877...24Z}, WISE 1541 -  \citealt{2014ApJ...783...68B,2017ApJ...842..118L,2019ApJ...877...24Z},WISE 0855 -  \citealt{2016AJ....152...78L, 2018ApJ...858...97M}. The mean Sonora Bobcat evolution model-derived temperatures for WISE 2056 and WISE 1541 are each 90 K and 60 K warmer than published BT Settl model fits published in \cite{2014ApJ...783...68B}, but consistent with the Morley model derived temperatures in the same paper.

\begin{turnpage}
\begin{deluxetable*}{lccccccc}
\footnotesize
\tablecaption{Photometric Information Used for Temperature Fitting}
\tablehead{
\colhead{\textbf{Object}} & \colhead{\textbf{Gl 570 D}} & \colhead{\textbf{2M 0415}} & \colhead{\textbf{UGPS 0722}} & \colhead{\textbf{WISE 0313}} &  \colhead{\textbf{WISE 2056}} &  \colhead{\textbf{WISE 1541}} & \colhead{\textbf{WISE 0855}}
}
\startdata
\textbf{Parallax} & 171.22 +/- 0.94 (1) & 175.2 +/- 1.7 (1) & 242.8 +/- 2.40 (2) & 134.3 +/- 3.6 (3) & 138.3 +/- 2.2 (3) & 167.1 +/- 2.3 (3) & 438.9 +/- 3.0 (3)\\
\hline\\
\textbf{Filter} & \multicolumn{6}{c}{\textbf{Magnitudes}}\\
\hline 
\textbf{2MASS J}    &    15.324 +/- 0.046 (4)   &   15.695 +/- 0.057 (4)   &   16.489 +/- 0.128 (4)   &   17.65 +/- 0.07 (5)   &   -    &   -  & - \\   
\textbf{2MASS H}    &   15.268 +/- 0.089 (4)   &   15.537 +/- 0.113 (4)   &   16.147 +/- 0.205 (4)   &   17.63 +/- 0.06 (5)   &   -    &   -  & -  \\   
\textbf{2MASS K$_{s}$}    &   15.242 +/- 0.156 (4)   &   15.429 +/- 0.201 (4)   &   $>$14.823 (4)   &   -    &   -    &   -  & -  \\   
\textbf{Y MKO}    &   -    &   -    &   17.37 +/- 0.02 (6)   &   18.27 +/- 0.05 (5)   &   19.94 +/- 0.05 (7)   &   21.63 +/- 0.13 (8) & -   \\   
\textbf{J MKO}    &   14.82 +/- 0.05 (9)   &   15.32 +/- 0.03 (10)   &   16.52 +/- 0.02 (6)   &   -    &   19.43 +/- 0.04 (7)   &   21.12 +/- 0.06 (7) & -   \\   
\textbf{H MKO}    &   15.28 +/- 0.05 (9)   &   15.70 +/-  0.03 (10)   &   16.90 +/- 0.02 (6)   &   -    &   19.96 +/- 0.04 (7)   &   21.07 +/- 0.07 (11) & -   \\   
\textbf{K MKO}    &   15.52 +/- 0.05 (9)   &   15.83 +/- 0.03 (10)   &   17.07 +/- 0.08 (6)   &   -    &   20.01 +/- 0.06  (7)   &   21.7 +/- 0.2 (11)  & -   \\   
\textbf{F105W}    &   -    &   -    &   -    &   -    &   -    &   22.204 +/- 0.044 (8) & 27.33 +/-  .19 (17)  \\   
\textbf{F110W}    &   -    &   -    &   -    &   -    &   -    &   -  & 26.00 +/- .12 (17)   \\   
\textbf{F125W}    &   -    &   -    &   -    &   -    &   -    &   21.871 +/- 0.023 (8) & 26.41 +/- .27 (18)   \\   
\textbf{F127M}    &   -    &   -    &   -    &   -    &   -    &   - & 24.36 +/- .09 (17)   \\   
\textbf{F140W}    &   -    &   -    &   -    &   -    &   19.524 +/- 0.007 (8)   &   -  & -    \\   
\textbf{F160W}    &   -    &   -    &   -    &   -    &   -    &   - & 23.86 +/- 0.03  (18)  \\   
\textbf{L$^{\prime}$}    &   12.98 +/- 0.05 (9)   &   13.28 +/- 0.05 (12)   &   13.4 +/- 0.3 (6)   &   -    &   -    &   -  & -   \\   
\textbf{M$^{\prime}$}    &   -    &   12.82 +/- 0.15 (12)   &   -    &   -    &   14.00 +/- 0.15 (13)   &   -  & -    \\   
\textbf{N}    &   -    &   -    &   10.28 +/- 0.24 (6)   &   -    &   -    &   -  & -   \\   
\textbf{IRAC1}    &   13.80 +/- 0.04 (14)   &   14.10 +/- 0.03 (14)   &   14.28 +/- 0.05 (6)   &   15.31 +/- 0.025 (15)   &   16.036 +/- 0.030 (5)   &   16.92 +/- 0.02 (7) & 17.28 +/- .02 (20)    \\   
\textbf{IRAC2}    &   12.12 +/- 0.02 (14)   &   12.29 +/- 0.02  (14)   &   12.19 +/- 0.04 (6)   &   13.268 +/- 0.017 (15)   &   13.924 +/- 0.018 (5)   &   14.12 +/- 0.01 (7)  & 13.88 +/- 0.02   \\   
\textbf{IRAC3}    &   12.77 +/- 0.11 (14)   &   12.87 +/- 0.07  (14)   &   -    &   -    &   -    &   -  & -   \\   
\textbf{IRAC4}    &   11.97 +/- 0.07 (14)   &   12.11 +/- 0.05  (14)   &   -    &   -    &   -    &   -  & -   \\   
\textbf{W1}    &   14.824 +/- 0.034 (16)   &   15.108 +/- 0.041 (16)   &   15.250 +/- 0.045 (16)   &   15.953 +/- 0.045 (16)   &   16.48 +/- 0.075 (16)   &   16.736 +/- 0.165 (16) & 17.819 +/- 0.327 (19)    \\   
\textbf{W2}    &   12.114 +/- 0.023 (16)   &   12.261 +/- 0.026 (16)   &   12.200 +/- 0.023 (16)   &   13.263 +/- 0.026 (16)   &   13.839 +/- 0.037 (16)   &   14.246 +/- 0.063 (16) & 14.016 +/- 0.048 (19)   \\   
\textbf{W3}    &   10.863 +/- 0.082 (16)   &   11.132 +/- 0.113 (16)   &   10.206 +/- 0.069 (16)   &   12.045 +/- 0.264 (16)   &   11.731 +/- 0.249 (16)   &   $>$12.2 (16) & 11.9 +/- 0.3 (13)  \\   
\textbf{W4}    &   $>$9.190 (16)   &   $>$8.638 (16)   &   $>$8.763 (16)   &   $>$8.668 (16)   &   $>$8.493 (16)   &   $>$8.892 (16)  & -  
\enddata
\tablecomments{Published infrared photometry and parallaxes used to fit for effective temperatures of objects in our sample. The references for each measurement are indicated by a number enclosed in parenthesis following the error value. While UGPS 0722 and WISE 0313 are close in adopted temperature, WISE 0313 has less photometric data to constrain models. The references associated with each number are: 
1 - \cite{2012ApJS..201...19D}, 2 - \cite{2012ApJ...753..156K},  3 - \cite{2019ApJS..240...19K}, 4 - \cite{2006AJ....131.1163S}, 5 - \cite{2011ApJS..197...19K}, 6 - \cite{2010MNRAS.408L..56L}, 7 - \cite{2013ApJ...763..130L}, 8 - \cite{2015ApJ...804...92S}, 9 - \cite{2001ApJ...556..373G}, 10 - \cite{2004AJ....127.3553K}, 11 - \cite{2015ApJ...799...37L}, 12 - \cite{2004AJ....127.3516G}, 13 - \cite{2017ApJ...842..118L}, 14 - \cite{2006ApJ...651..502P}, 15 - \cite{2012ApJ...753..156K}, 16 - \cite{2013yCat.2328....0C}(AllWISE), 17 - \cite{2016AJ....152...78L}, 18 - \cite{2016ApJ...823L..35S}, 19 - \cite{2014AJ....148...82W}, 20 - \cite{2016ApJ...832...58E}.
\label{tbl:published-photometry}}
\end{deluxetable*}
\end{turnpage}

\begin{deluxetable*}{lcccccc}
\tablecaption{Sample of Substellar Objects with M band Spectra}
\tablecomments{Objects in our analysis which have temperatures from 75 K to 125 K and $M$-band spectral observations. WISE 0855's spectral type was inferred with photometry in \cite{2016ApJ...823L..35S}, no near infrared spectral observations have been taken. The range of possible surface gravities for brown dwarfs in the last column are from the cloudless Sonora Bobcat evolution models and assume ages between 1 Gyr and 10 Gyr.}
\tabletypesize{\footnotesize} 
\tablehead{
\colhead{Object} &  \colhead{Adopted}          & \colhead{Effective Temperature}   & \colhead{Temperature}           & \colhead{Spectral} & \colhead{Spectral Type} & \colhead{Surface Gravity}   \\
\colhead{}       &  \colhead{ Temperature (K)} & \colhead{Range (K)}              & \colhead{Reference}             & \colhead{Type}     &  \colhead{Reference}         & \colhead{ Range log(cm s$^{-2}$)}}
\startdata
Gl 570 D           & 750 K  & 716 - 812 & re-fit in this work       & T7.5           & \cite{2006ApJ...637.1067B} & 4.7 - 5.4 \\
2MASS J0415-0935   & 700 K  & 649 - 734 & re-fit in this work       & T8             & \cite{2006ApJ...637.1067B} & 4.6 - 5.3  \\
WISE 0313          & 650 K  & 606 - 685 & re-fit in this work       & T8.5           & \cite{2011ApJS..197...19K} & 4.6 - 5.3 \\
UGPS 0722          & 550 K  & 522 - 558 & re-fit in this work       & T9             & \cite{2011ApJ...743...50C} & 3.7 - 4.4  \\
WISE 2056          & 500 K  & 471 - 522 & re-fit in this work       & Y0             & \cite{2015ApJ...804...92S} & 4.4 - 5.0 \\
WISE 1541          & 400 K  & 396 - 434 & re-fit in this work       & Y1             & \cite{2015ApJ...804...92S} & 4.3 - 4.9 \\
WISE 0855          & 250 K  & 249 - 260 & re-fit in this work       & $>$Y4          &  \cite{2019ApJS..240...19K} & 3.5 - 4.5  \\
Jupiter            & 126 K  & - & \cite{2012JGRE..11711002L}        & -              & - & 3.4
\label{tbl:objects}
\end{deluxetable*}

\subsection{M-Band Spectral Sequence}

\begin{figure*}
\centering
\includegraphics[width=3in]{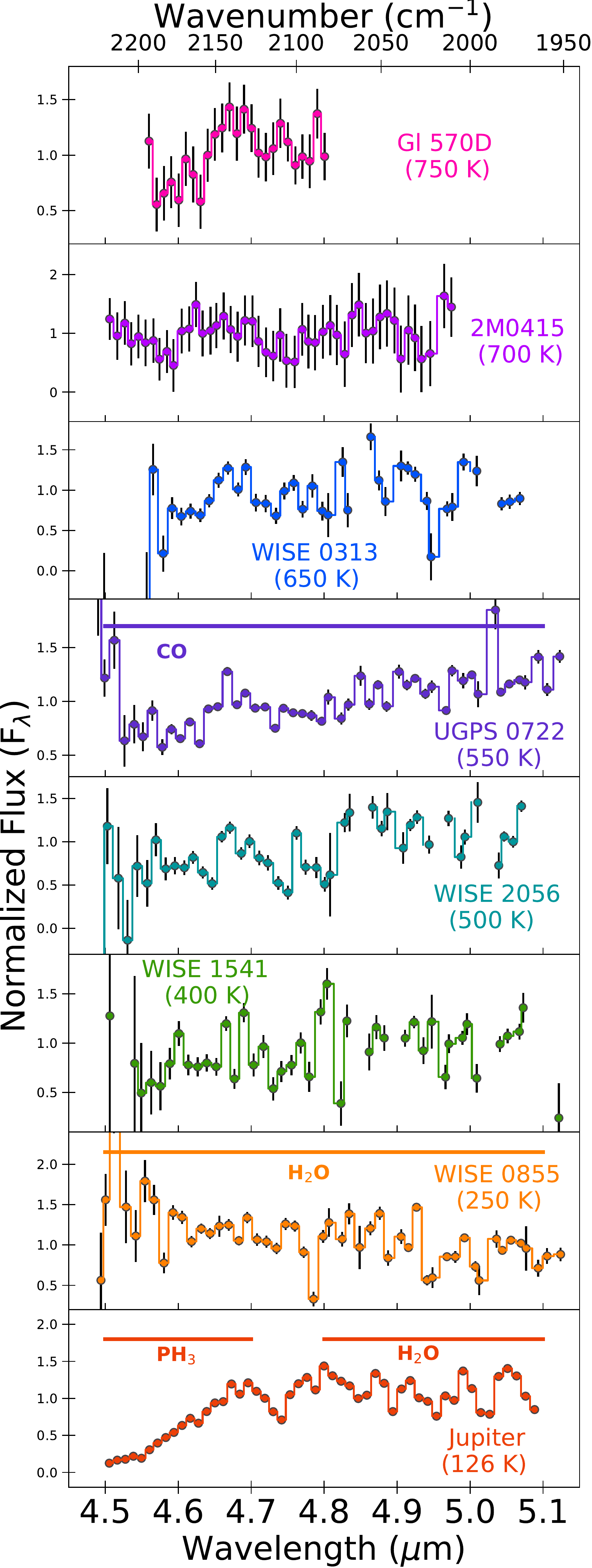}
\caption{Normalized $M$-band spectra of cool substellar objects. The y-axis of each plot are set differently to emphasize carbon monoxide absorption across the spectra. All of the brown dwarfs show evidence of carbon monoxide absorption indicating that their atmospheres are out of chemical equilibrium. Carbon monoxide has been detected in Jupiter at very high spectral resolution. \citep{2002Icar..159...95B}.}
\label{fig:m_band_sequence}
\end{figure*}

\begin{figure}
\centering
\includegraphics[width=3.3in]{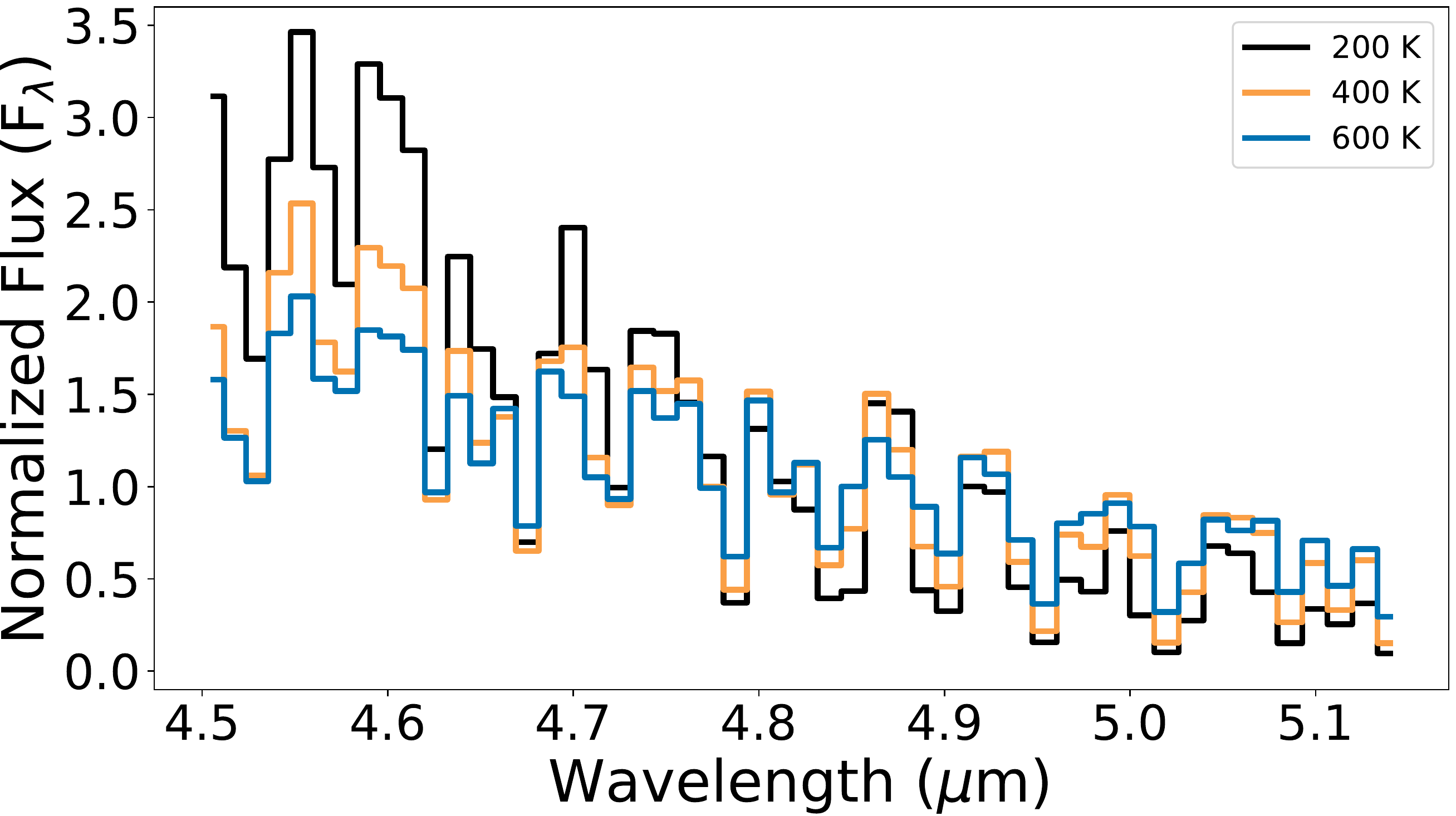}
\caption{Normalized $M$ band spectra of cloudless, solar metallicitiy brown dwarfs of varying temperatures. In normalized space, most of the temperature change can be seen shortward of 4.75 $\mu$m and at lower temperatures the spectral slope becomes steeper across the $M$-band.}
\label{fig:temp-change}
\end{figure}

\begin{figure}
\centering
\includegraphics[width=3.3in]{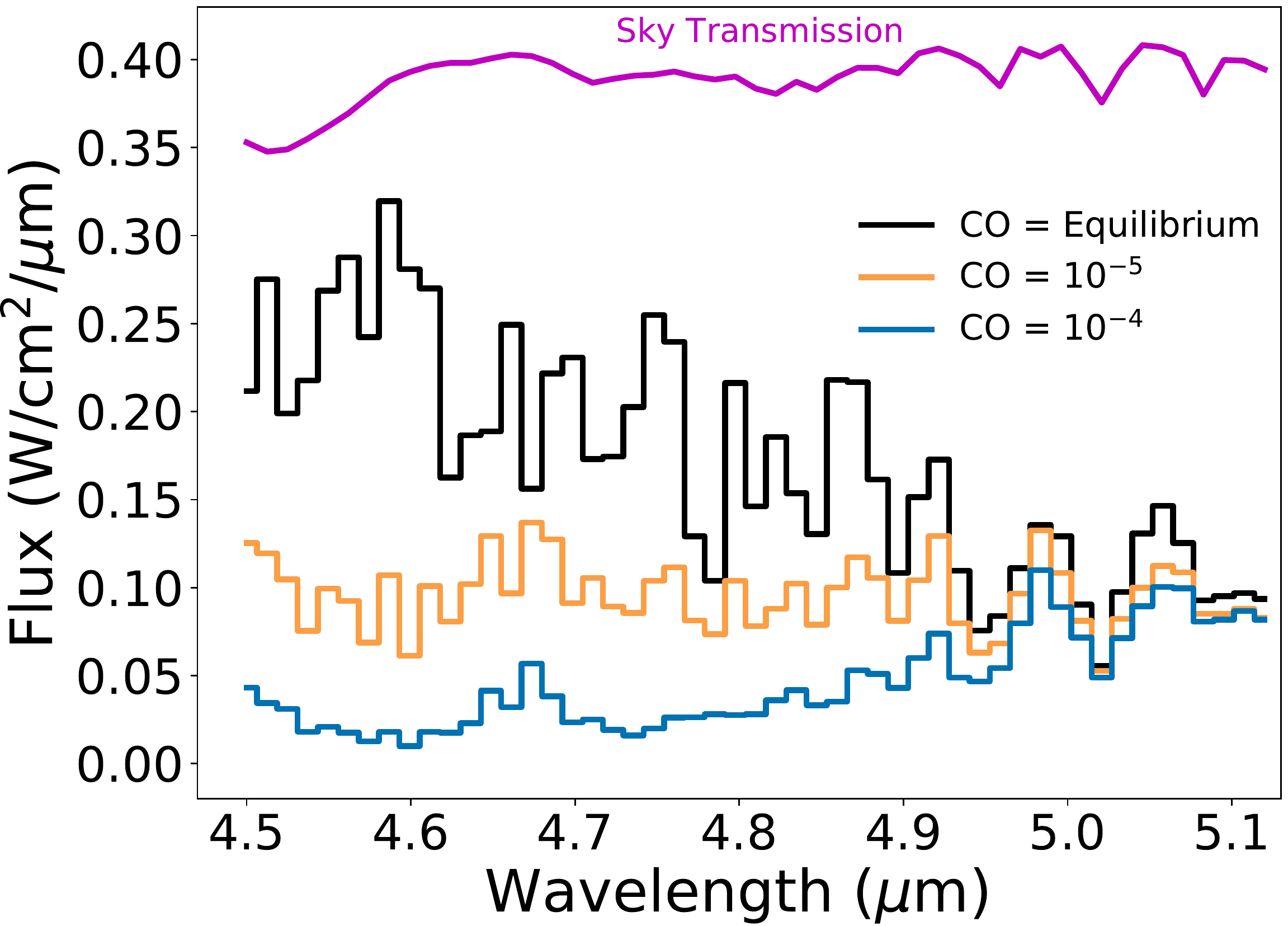}
\caption{$M$-band spectra of a 550 $K$ cloudless, solar metallicity brown dwarf with varying mole fractions of carbon monoxide. The carbon monoxide abundance influences the entire $M$-band spectral region. The sky transmission is plotted in magenta at the top of the figure and lower transmission areas typically correspond to higher telluric errors.}
\label{fig:co-change}
\end{figure}

Figure~\ref{fig:m_band_sequence} shows the $M$-band spectra of brown dwarfs with effective temperatures from 750 K to 250 K and Jupiter. The $M$-band spectrum of 2M0415 was taken with the AKARI spacecraft and published in \cite{2012ApJ...760..151S}. The adopted $M$-band spectrum for Gl570D was taken using  Gemini/NIRI and published in \cite{2009ApJ...695..844G}. The AKARI spectrum of Gl570D from \cite{2012ApJ...760..151S} was not used because it has lower signal-to-noise than the Gemin/NIRI spectrum. The spectrum of WISE 0855 is the re-reduced version from this work. Jupiter's spectrum taken with the Short-Wavelength Spectrometer (SWS) on ISO \citep{1996A&A...315L.397E} was binned down from a resolution of 31,000 to 370. WISE 0855 has a distinct spectral slope compared to the rest of the sample that closely resembles an equilibrium atmosphere dominated by water across the $M$-band as shown in Figure~\ref{fig:temp-change}. UGPS 0722 is the object with the highest signal-to-noise in our program and shows an absorption feature across 4.5 $\mu$m to 4.8 $\mu$m, peaking at $\sim$4.7 $\mu$m that looks similar to carbon monoxide which has been seen in T-dwarfs like Gl229B (\cite{1997ApJ...489L..87N}, Figure~\ref{fig:co-change}). Gl 570D, 2M0415, and WISE 0313 show a similar absorption feature as UGPS 0722, but with less data quality for WISE 0313 and a flatter spectral slope for 2M0415. Gl 570D's absorption feature and spectral slope closely resemble UGPS 0722's spectrum\footnote{We adopt the Gemini spectrum because it has higher signal-to-noise than the AKARI data. The AKARI spectrum does not have an obvious CO feature, but it is consistent within error to the Gemini/NIRI spectrum.}. WISE 2056 and WISE 1541 look relatively flat, but appear to have increasing flux at redder wavelengths. Carbon monoxide and the shape of its $M$-band feature are discussed in more detail in Section \ref{subsec:non-eq co}. WISE 1541 (400 K), WISE 0855 (250 K), and Jupiter (126 K) cover a temperature range of less than 300 K, yet they express different spectral shapes and molecular absorption features. Jupiter's phosphine (\ce{PH3}) feature that stretches from 4.5 $\mu$m to 4.7 $\mu$m is a tracer of atmospheric quenching, but \ce{PH3} is not obvious for any of the brown dwarfs. However, Jupiter's metallicity abundance is 3 times the solar value and that is important when placing it in context to brown dwarfs which are being modeled and discussed in this paper.

\section{Model Comparisons}
\label{sec:model_comparison}

\subsection{Equilibrium Models}
\begin{figure}
\centering
\includegraphics[width=3.3in]{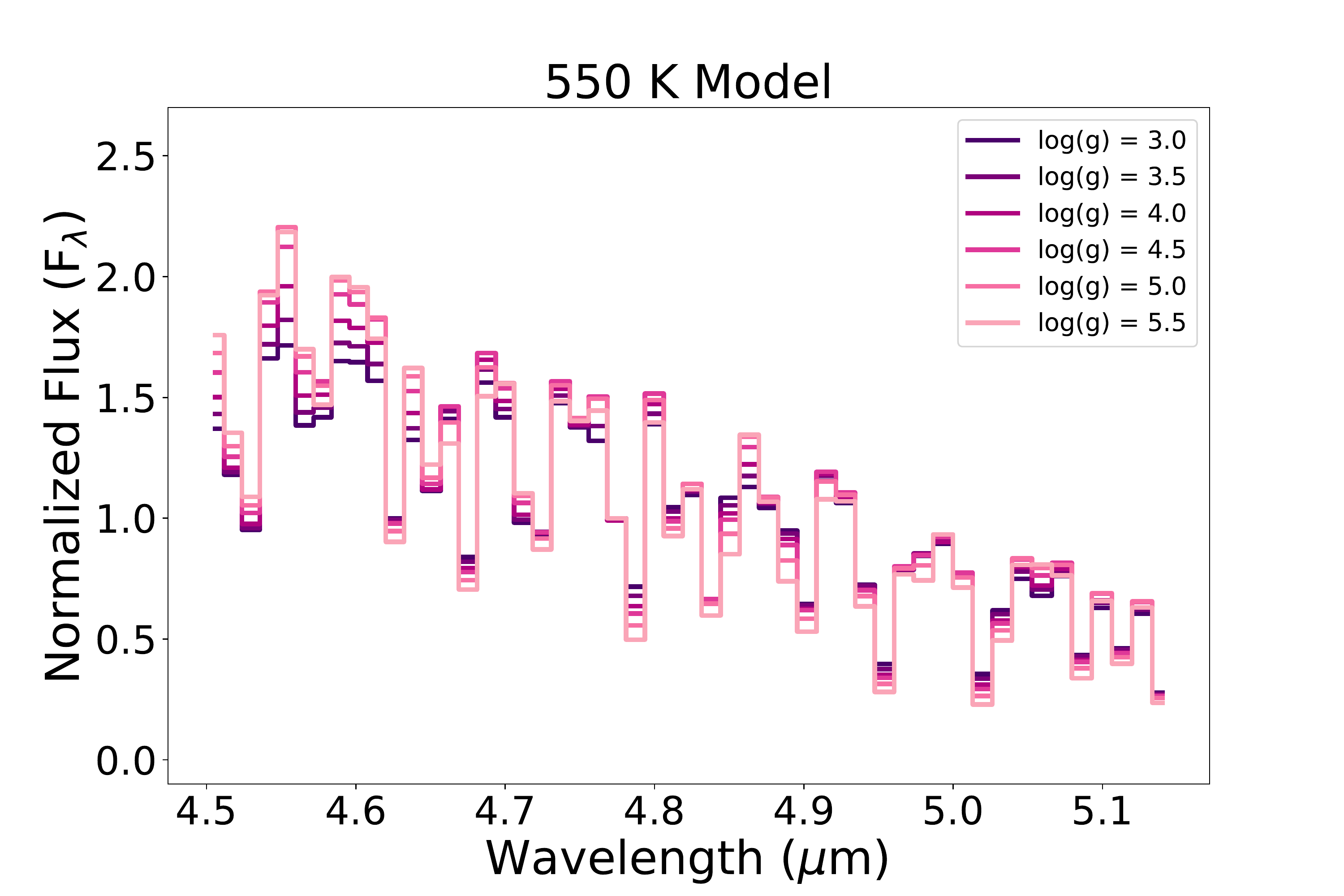}
\caption{A 550 K, cloudless, solar metallicity model with a range of surface gravities plotted. Where surface gravity influences the spectrum the most, the data quality is typically poor(see Figure~\ref{fig:co-change}). }
\label{fig:gravity-change}
\end{figure}

\begin{figure}
\centering
\includegraphics[width=3.3in]{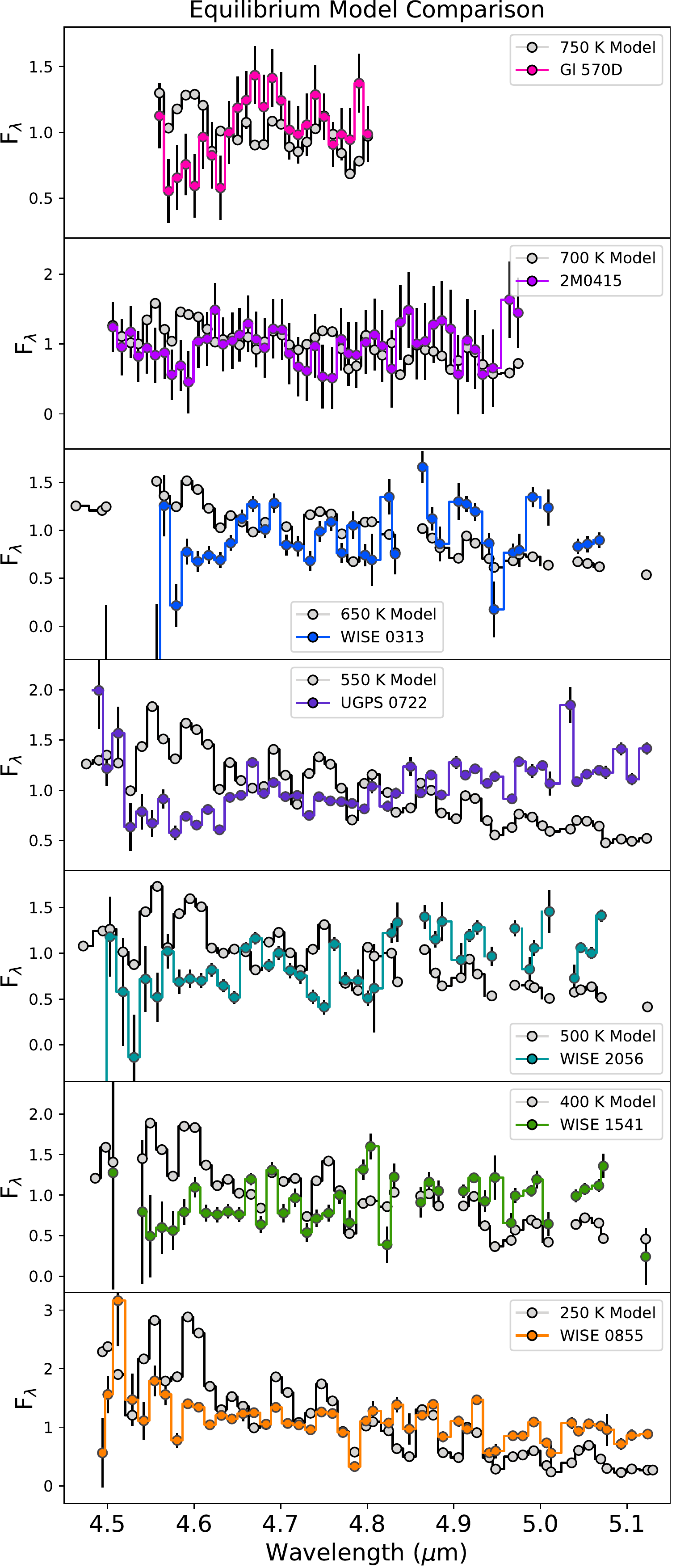}
\caption{The $M$-band spectra of each brown dwarf (Color) plotted against the same temperature cloudless, solar metallicity, chemical equilibrium model (Light Grey). The models are binned down to the number of elements in the data and then interpolated onto the data's spectral grid. In all cases the models have steeper spectral slopes, which cannot be explained by temperature alone. The y-axis does not show the full range of spectral data to emphasize absorption features.}
\label{fig:eq-fit}
\end{figure}

In equilibrium, the  dominant absorber across the $M$-band is water for brown dwarfs below effective temperatures of 800 K and the spectral slope is most sensitive to cloudiness and temperature (\cite{2014ApJ...787...78M, 2018ApJ...858...97M}, see Figure~\ref{fig:temp-change}). A model comparison of the cloudless, equilibrium case for each brown dwarf is done using the adopted temperatures found in Section~\ref{sec:sequence} (Figure~\ref{fig:eq-fit}), assuming surface gravity of log(g) = 4.5. The surface gravities are not known, but have very little effect on the shape of the normalized $M$-band spectra (Figure ~\ref{fig:gravity-change}). The cloudless equilibrium models do not adequately fit the spectra of the brown dwarfs, because of increased flux on the blue side of the models relative to the data. However, water vapor absorption beyond $\sim$4.85 $\mu$m in the equilibrium models do line up with a few dips in the data for some brown dwarfs. 

\subsection{Disequilibrium, Carbon Monoxide Enhanced Models}
\label{subsec:non-eq co}
Vertical atmospheric mixing can bring up carbon monoxide-rich gas from higher temperature regions into pressure levels probed by the observations. CO is usually observed in the $K$-band spectra of L and early T dwarfs, but the much stronger transitions of the fundamental band, centered near 4.7 $\mu$m, can be observed in the $M$ band in late T dwarfs, when the $K$-band CO features are obscured by \ce{CH4} absorption.  As mentioned earlier, the brown dwarf Gl229B (900 K) showed absorption across the $M$-band (Figure 1 and 2 in \cite{1997ApJ...489L..87N}) that is best-fit using atmospheric models with enhanced abundances of carbon monoxide. CO has also been previously detected in the atmosphere of Jupiter \citep{2002Icar..159...95B}. UGPS 0722 and other brown dwarfs in our sample share similar characteristics with the Gl229B spectrum, therefore we try to understand how much carbon monoxide can explain the variations between the spectra in our sample.

\begin{figure}
\centering
\includegraphics[width=3.3in]{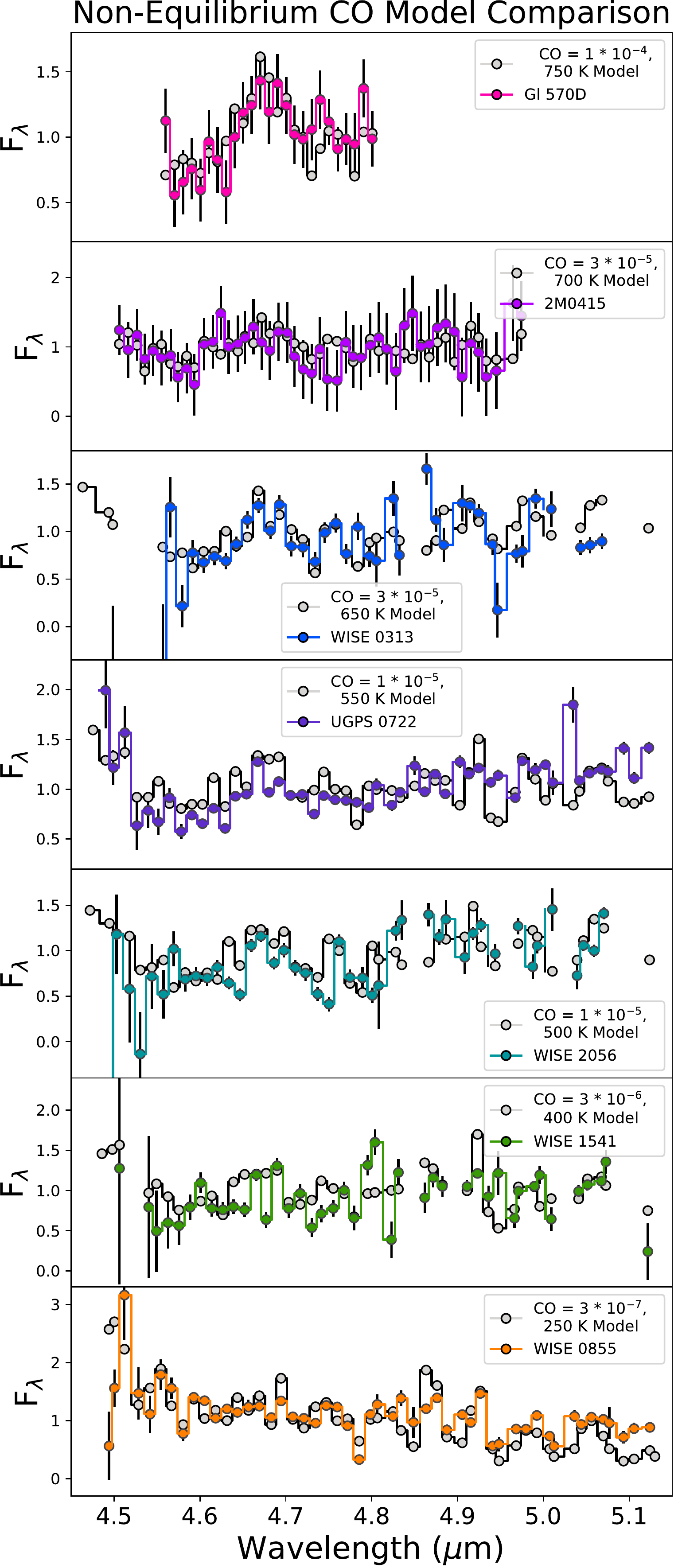}
\caption{The $M$-band spectra of each brown dwarf (Color) plotted with a cloudless, solar metallicity model with the best fit adjusted carbon monoxide mole fraction (Light Grey). All brown dwarfs need enhanced abundances of CO for a better fit. Warmer objects typically have more carbon monoxide. The y-axis does not show the full range of spectral data to emphasize absorption features}
\label{fig:non-eq-fit}
\end{figure}

The carbon monoxide enhanced models for the normalized $M$-band spectra fitting use pressure-temperature profiles from the Sonora Bobcat models. The carbon monoxide abundance profile is changed for different quench abundances of carbon monoxide. Moderate resolution spectra are generated as described in the Appendix of \cite{2015ApJ...815..110M}. For each adopted effective temperature and surface gravity of log(g) = 4.5, a model grid with \ce{CO} mole fractions between 10$^{-7}$ and 10$^{-4}$ (spaced by a factor of 3) are fit against the $M$-band spectra. Adding carbon monoxide alters the $M$-band from 4.5 $\mu$m to 4.95 $\mu$m by flattening out regions of the spectrum and leaving a peak at 4.7 $\mu$m (Figure~\ref{fig:co-change}). Figure~\ref{fig:non-eq-fit} shows the best fit carbon monoxide enhanced models for each brown dwarf. All of the brown dwarfs needed disequilibrium carbon monoxide abundances to achieve a better spectral fit. These best-fit mole fraction values of carbon monoxide range from values of 10$^{-4}$ to 10$^{-7}$, with lower abundances at lower effective temperature. WISE 1541 still has some discrepancy on the blue portion of the spectrum, where the disequilibrium model has a slightly steeper slope than the data. 

The expected equilibrium mole fraction of carbon monoxide falls rapidly from $\sim$10$^{-7}$ to $\sim$10$^{-18}$ \citep{2002Icar..155..393L} within the $M$-band photopshere for effective temperatures between 750 K and 500 K, which points to atmospheric quenching driving the CO abundances of our objects to disequilibrium. All of the objects in our sample have disequilibrium \ce{CO} abundances, therefore, atmospheric quenching may be ubiquitous among cool brown dwarfs and gas giant exoplanets. Directly imaged, gas giant exoplanets found around younger stars tend to have lower surface gravities relative to field brown dwarfs, making them prone to stronger atmospheric mixing and expressing disequilibrium abundances of molecules \citep{2014ApJ...797...41Z}. If future directly imaged gas giants frequently fall somewhere within the effective temperature range of the objects in Table~\ref{tbl:objects}, $M$-band spectroscopy will be vital for their characterization. The binned spectra presented in this work are fairly low resolution (R~$\sim$ 370), but Jupiter's CO abundance is only detectable with high resolution (ex. R $\sim$ 42,680 \citealt{2002Icar..159...95B}) ) spectroscopy. As we image colder and older gas giants, higher resolution modes will be essential for exoplanet-focused instruments.

\subsection{Modeling Clouds and CO in WISE 0855}

Carbon monoxide and clouds both have the potential to significantly alter the spectral shape across the $M$-band \citep{2014ApJ...787...78M} and we briefly explore their combined effect in this Section. The models described in Section 2.3 of \cite{2018ApJ...858...97M} that have homogeneous clouds and varying amounts of carbon monoxide are used to fit WISE 0855's spectrum. The effective temperature is fixed at 250 K and the surface gravity is fixed at log(g) equal to 4. The lower surface gravity used for WISE 0855 is from \cite{2018ApJ...858...97M}, though it has a minor effect on the spectrum. Our cloudy model grid covers CO mole fractions between 10$^{-5}$ and 10$^{-7.5}$, including a model with no CO. The clouds are parameterized by the sedimentation efficiency (f$_{\rm sed}$) defined in \cite{2001ApJ...556..872A}, where lower values of f$_{\rm sed}$ produce extended lower density clouds and higher values produce thinner, dense clouds. Cloudy models have f$_{\rm sed}$ values of 2 - 10, spaced by increments of 2. Each model is binned to the resolution of the data then interpolated onto the wavelength spacing of the data for fitting. 

\begin{figure}
\includegraphics[width=3.3in]{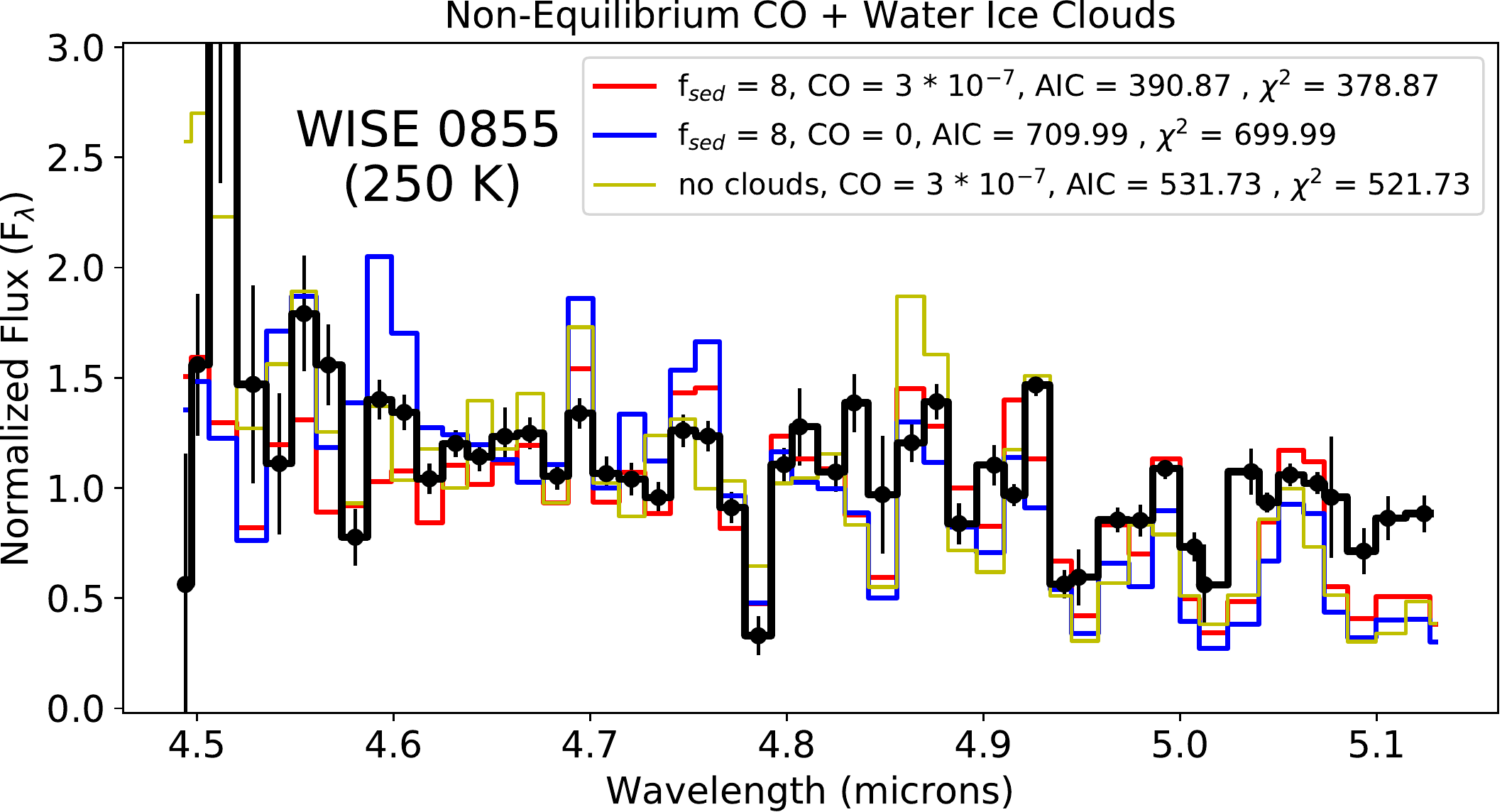}
\caption{Comparisons of models with clouds, CO, and clouds with CO. Both WISE 0855 and WISE 1541 are better fit with CO and clouds, WISE 1541 shows significant deviations from the models at shorter wavelengths within the $M$-band.}
\label{fig:co and clouds}
\end{figure}

The best fit model for WISE 0855 has an f$_{\rm sed}$ value of 8 and CO mole fraction of 10$^{-6.5}$ (Figure~\ref{fig:co and clouds}). According the Akaike information Criterion (AIC), adding clouds to the model is justified because the metric penalizes the addition of extra parameters while assessing relatively model quality. The AIC is chosen over the Bayesian information criterion (BIC) because the true model is not known (see Figure~\ref{fig:photometry}) and we are looking for a relative comparison of goodness of fit.  The AIC of the best-fit cloud and CO model is 390.87, while the AIC values of the cloud-only and CO-only models are 709.99 and 531.73 respectively. Clouds improve the spectral fits compared to equilibrium in all cases, but clouds typically make very broad wavelength changes along a spectrum and should be constrained with a full spectrum or spectral energy distribution of photometry in future work. Our analysis supports the prediction from \cite{2014ApJ...787...78M}, that water clouds are a significant opacity source in brown dwarfs under 400 K, and the clouds become optically thick at temperatures between 375 K - 350 K. 

The presence of water clouds in WISE 0855's atmosphere has been debated in the literature, but future space-based facilities like JWST may be able place better constraints on water clouds or other condensates. The first evidence for water clouds on WISE 0855 came from the work of \cite{2014ApJ...793L..16F}, which used J$_{MKO}$ - W2 colors to rule out atmospheric models that contained no clouds and models with only sulfide clouds. The first spectrum of WISE 0855 published in \cite{2016ApJ...826L..17S} showed that a grey opacity source, representing water clouds, produces a better spectral fit to the $M$-band data than an atmospheric model with no clouds. The same spectrum of WISE 0855 was later modeled with water ice clouds \citep{2018ApJ...858...97M} showing the same results as \cite{2016ApJ...826L..17S}, providing stronger evidence in favor of water clouds. The photometric monitoring done with \textit{Spitzer} presented in \cite{2016ApJ...832...58E} showed that WISE 0855's variability was irregular and did not have the same pattern between simultaneous \textit{Spitzer} [3.6] and \textit{Spitzer} [4.5] observations. The authors claimed that the irregular photometric series was similar to other T-dwarfs, therefore the source of the variability must arise from opacity sources common to both Y and T-dwarfs, which could not be water clouds. At the moment, it is likely not feasible to attempt low-resolution spectroscopic monitoring of WISE 0855 from the ground and near-infrared spectroscopic data have not been taken with the Hubble Space Telescope. Photometric measurements may not be able to distinguish cloud variability from heterogenous disequilibrium chemistry or hot spots, therefore spectrographs like NIRSpec on JWST will be needed for space-based, time series, observations to characterize WISE 0855 atmosphere.

\section{Atmospheric Quenching and Other Disequilibrium Molecules}
\label{sec:atm_mixing}

\subsection{Inferred Atmospheric Mixing from CO}

The chemistry of cool substellar atmospheres is complex and relies on many poorly constrained values such as metallicity, gravity, abundances of trace gases, and vertical mixing rates in the atmosphere. In this section we are not attempting to provide exact calculations to explain the disequilibrium abundances of carbon monoxide in our spectra, but understand what assumptions can reasonably explain the spectral features within our sample and if fundamental assumptions need to be changed in future work.

The strength of large scale vertical mixing is often parameterized by the eddy diffusion coefficient (K$_{zz}$) which is defined by the equation
\begin{equation}
    K_{zz} = \frac{L^{2}}{\tau_{\rm mix}}
\label{eq:eddy}
\end{equation}
where $L$ is a length scale of mixing and $\tau_{\rm mix}$ is the timescale of mixing. Higher values of K$_{zz}$ correspond to shorter mixing timescales ($\tau_{\rm mix}$) within an atmosphere. Molecules have a chemical timescale ($\tau_{\rm chem}$) at which they are either created and/or turned into new products and these processes are often strongly dependent on temperature and pressure. If at a given pressure and temperature, the mixing timescale of an atmosphere is greater than the chemical timescale of a net reaction, that gas has sufficient time to reach \textbf{chemical equilibrium}. If the chemical timescale of a net reaction is greater than the rate of mixing at specific pressure and temperature, the molecule will \textbf{not be in chemical equilibrium}. In the disequilibrium case, it is often assumed that the detected abundance of a molecular species in the upper atmosphere is set by the \textbf{quench point}, where the chemical timescale is equal to the mixing timescale.

Following the prescription of \cite{2014ApJ...797...41Z}, the length scale is assumed to be equal to the pressure scale height ($H$), which is calculated by
\begin{equation}
    H = L = \frac{k_{b}T}{\mu m_{h} g}
\label{eq:scale height}
\end{equation}
Assuming a log(g) equal to 4.5, the median scale height for the 250 K - 750 K effective temperature range calculated between pressures of 10$^{-4}$ to 10$^{5}$ bars  spans from 32 km to 61 km. The corresponding mixing timescale along the pressure-temperature profile is 
\begin{equation}
    \tau_{\rm mix} = \frac{H^{2}}{K_{zz}}
\label{eq:tmix}
\end{equation}
The eddy diffusion coefficient value of an object is estimated by finding the point along the pressure-temperature profile where the best-fit molecular abundance is found in chemical equilibrium at solar metallicity, then finding the $K_{zz}$ value required to match the mixing timescale to the chemical timescale at that point. We explore the disequilibrium chemistry of three molecules: carbon monoxide (\ce{CO}), phosphine (\ce{PH3}), and ammonia (\ce{NH3}). The equilibrium abundance information as a function of pressure and temperature for \ce{CO} and \ce{NH3} is taken from the Sonora Bobcat structure models. The abundance of \ce{PH3} along a pressure-temperature profile is calculated by doing 2-D linear interpolation\footnote{scipy.interpolate.griddata} over the mole fraction contour lines versus temperature and pressure in \cite{2006ApJ...648.1181V}. The chemical timescale equations for \ce{CO} and \ce{NH3} are from Equations 12, 13, 14, and 32 of \cite{2014ApJ...797...41Z}, which are derived from empirical exponential fits to one dimensional models that treat vertical mixing like diffusion. The chemical timescales for \ce{PH3} are derived for each pressure-temperature profile using techniques in \cite{2006ApJ...648.1181V}. A constant entropy adiabat in the deep atmosphere is assumed to extend the pressure range of the Sonora grid from a log(P(bars)) of 2.2 out to a depth of log(P (bars)) of 6. The value of $\mu$ along the extended pressure-temperature profile is taken as the last value at the high pressure end of the original profile. The mean molecular weight only changes by .007\% over the entire profile before extension. With the extended pressure-temperature profiles and the appropriate timescale information we can estimate the K$_{zz}$ values and  quench points of \ce{CO}, \ce{PH3}, and \ce{NH3} for each brown dwarf in our sample and Jupiter. 

\begin{figure*}
\centering
\includegraphics[width=6in]{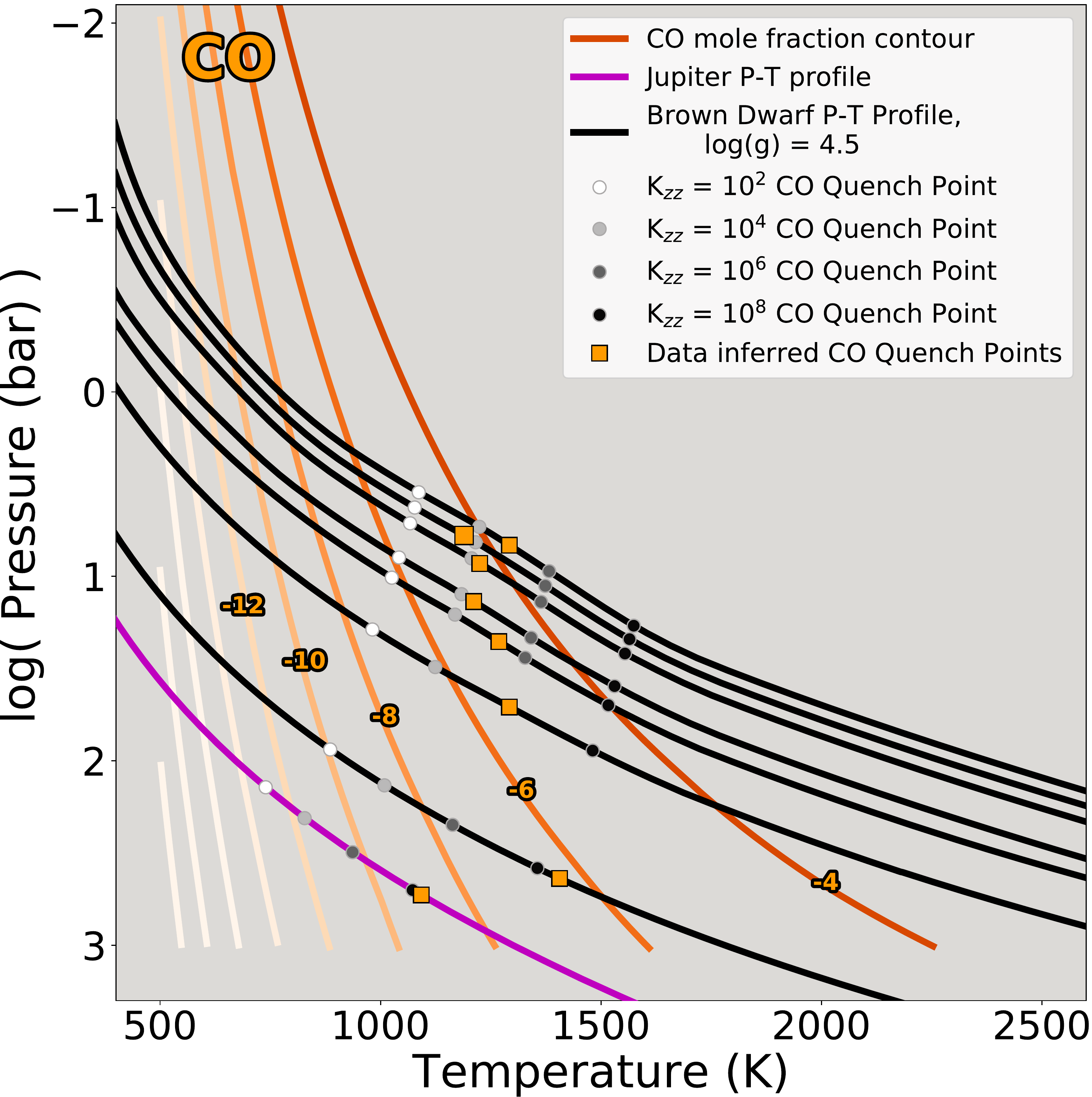}
\caption{This log pressure vs temperature plot contains contour lines of carbon monoxide (CO) mole fractions (orange lines) under equilibrium conditions. The numbers on each orange line represent the exponent value of that mole fraction contour line. Hotter temperatures and pressures correspond to higher abundances of CO. The black lines are pressure-temperature profiles of the adopted log(g) = 4.5 model for each brown dwarf in our sample. From top to bottom the effective temperatures are 750 K (Gl 570D) , 700 K (2M0415), 650 K (WISE 0313), 550 K (UGPS 0722),  500 K (WISE 2056), 400 K (WISE 1541), and 250 K (WISE 0855). Jupiter's pressure-temperature profile (magenta) is created using a hydrogen-helium mixture equation of state form \citep{2019ApJ...872...51C} and structure modeling developed in \cite{2016ApJ...831...64T} . Jupiter's P-T profile is below all of the brown dwarf pressure-temperature profiles. The greyscale dots represent quench points for fixed values of $K_{zz}$. The orange squares are the quench points based on the best fit carbon monoxide enhanced model. Jupiter has a quench point estimate using CO constraints from \cite{2002Icar..159...95B}. The best fit CO mole fraction tends to decrease at lower effective temperatures, but the amount of mixing required to keep those disequilibrium abundances increases towards lower effective temperatures. Jupiter has a lower surface gravity than the adopted brown dwarfs causing the offset in constant $K_{zz}$ value quench points. }
\label{fig:quench-co}
\end{figure*}

The atmospheric quench points are calculated for each brown dwarf and Jupiter using the best fit CO values. Initially, we assume a log(g) = 4.5 for all brown dwarfs and Jupiter has a log(g) of 3.4. Jupiter's \ce{CO} abundance profile is influenced by external factors, but the pressures probed by the $M$-band extend to the depth where \ce{CO} is quenched \citep{2002Icar..159...95B, 2010Icar..209..602V, 2011ApJ...738...72V}. In Figure~\ref{fig:quench-co} the \ce{CO} inferred quench points are plotted along pressure-temperature profiles representative of our gas giant sample. The quench points for fixed values of K$_{zz}$ are also plotted. Cooler objects have less CO, but require larger eddy diffusion coefficients to maintain those values. For fixed surface gravity and $K_{zz}$ values, warmer brown dwarfs will have more \ce{CO} gas detected in the atmosphere. Lower gravity objects will have higher abundances of CO for the same K$_{zz}$ value at fixed effective temperature. Lower gravity caused by youth is one of the reasons why directly imaged exoplanets near the L to T transition can display enhanced CO relative to methane dominated T-dwarfs of similar effective temperature \citep{2011ApJ...735L..39B,2018ApJ...869...18M}.

CO is a promising tracer for atmospheric quenching in gas giants, but different spectral resolutions will be required to capture the full range of quench levels possible. Between $K_{zz}$ values of $10^{2}$ cm$^{2}$ s$^{-1}$ and $10^{8}$ cm$^{2}$ s$^{-1}$  the CO mole fraction changes by a factor of 1000 at fixed effective temperature. From 700 K to Jupiter (126 K) the range of CO abundances spans mole fractions of  10$^{-13}$ to 10$^{-4}$. Jupiter's relatively low CO abundance has only been detected through high resolution spectroscopy, therefore brown dwarfs with smaller $K_{zz}$ values will also need more resolution at sufficiently high signal-to-noise for CO detections. The best fit CO abundances, inferred K$_{zz}$ values, \ce{PH3} abundances, and \ce{NH3} abundances are listed in Table~\ref{tbl:quench abundances}. The abundances of  \ce{PH3} and \ce{NH3} based on \ce{CO} will be discussed in Sections~\ref{sec:PH3} and~\ref{sec:NH3}.

\begin{figure*}
\centering
\includegraphics[width=6in]{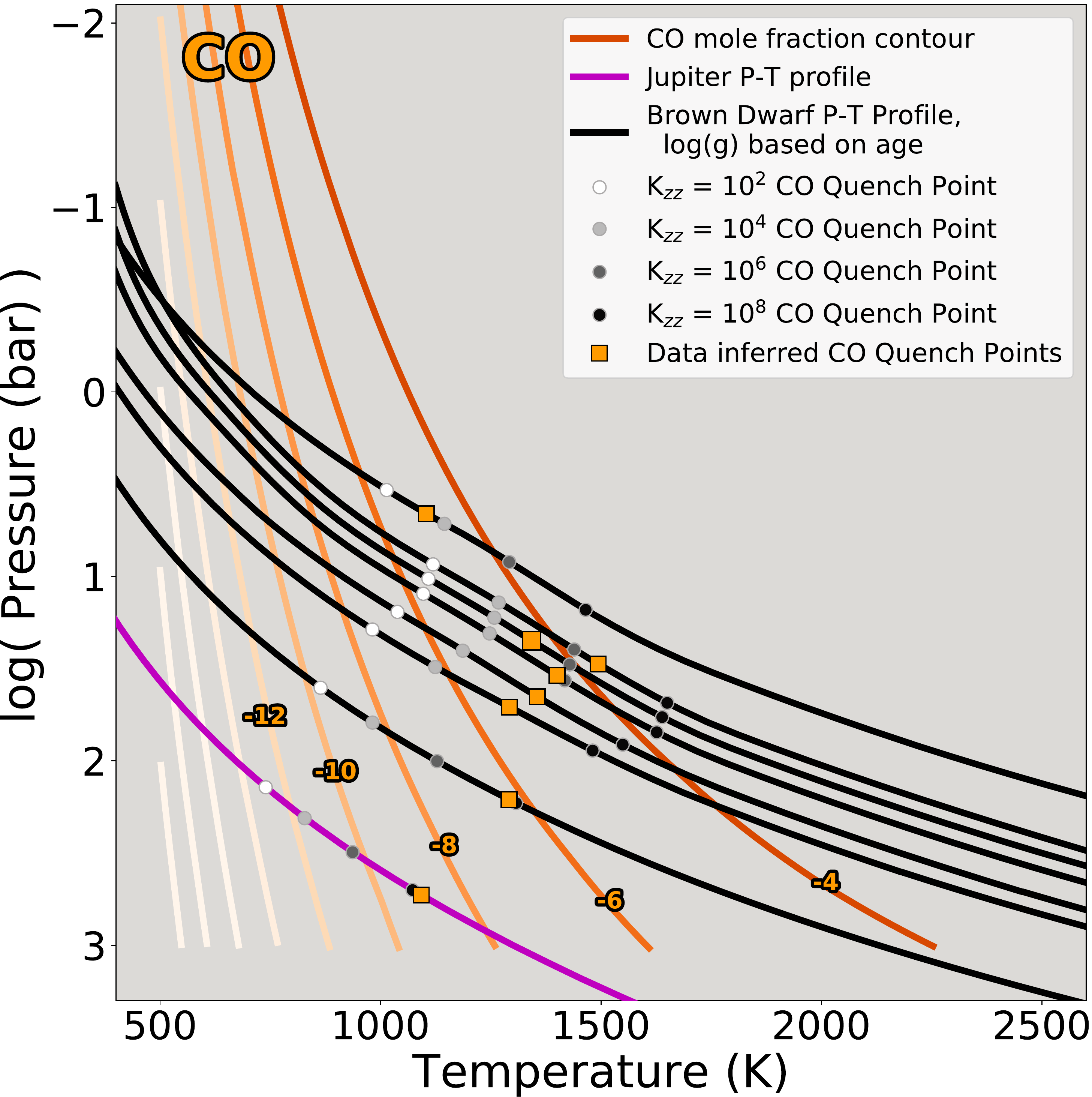}
\caption{Similar labelling as in Figure~\ref{fig:quench-co}. The pressure versus temperature profiles for each brown dwarf, but with different surface gravities based on age (See Table~\ref{tbl:objects}.) are plotted instead. UGPS 0722 is the upper most P-T profile, due to its lower surface gravity log(g) = 4.  The following P-T profiles from top to bottom are Gl 570D (log(g) = 5) , 2M0415 (log(g) = 5), WISE0313 (log(g) = 5), WISE 2056 (log(g) = 4.75),  WISE 1541 (log(g) = 4.5),  WISE 0855 (log(g) = 4).}
\label{fig:quench-co-age}
\end{figure*}

The surface gravities of the brown dwarfs in our sample have not been directly measured, but it is an important parameter for inferring K$_{zz}$ from the data. Warmer brown dwarfs have higher surface gravities for fixed age ranges, which leads to higher values of K$_{zz}$. However, when accounting for this effect Jupiter and WISE 0855 still have higher K$_{zz}$ relative to their warmer counterparts (Figure~\ref{fig:quench-co-age}). Since surface gravity is an unconstrained parameter, K$_{zz}$ is inferred for each surface gravity value that exists on the Sonora model grid described in Section~\ref{sec:model_comparison} for all of the brown dwarfs in our sample (Figure~\ref{fig:kzz-temp}).

\subsection{CO and Implications for Atmospheric Energy Transport}

 Brown dwarfs have deep convective interiors and radiative upper atmospheres, which respectively use convective motion and thermal radiation as the primary mode of energy transport (\citealt{2012sse..book.....K}, Figure \ref{fig:convection-zones}). Brown dwarfs with effective temperatures between 400 K and 750 K have detached upper convective zones with radiative zones in between that could inhibit transport of higher temperature, \ce{CO}-abundant material into the lower pressure regions of the atmosphere. The radiative zones in between the detached convective zones have steeper temperature gradients that could hinder convection and lower the value of  K$_{zz}$ estimated from the \ce{CO} abundance.
 
\begin{figure*}
\centering
\includegraphics[width=6.5in]{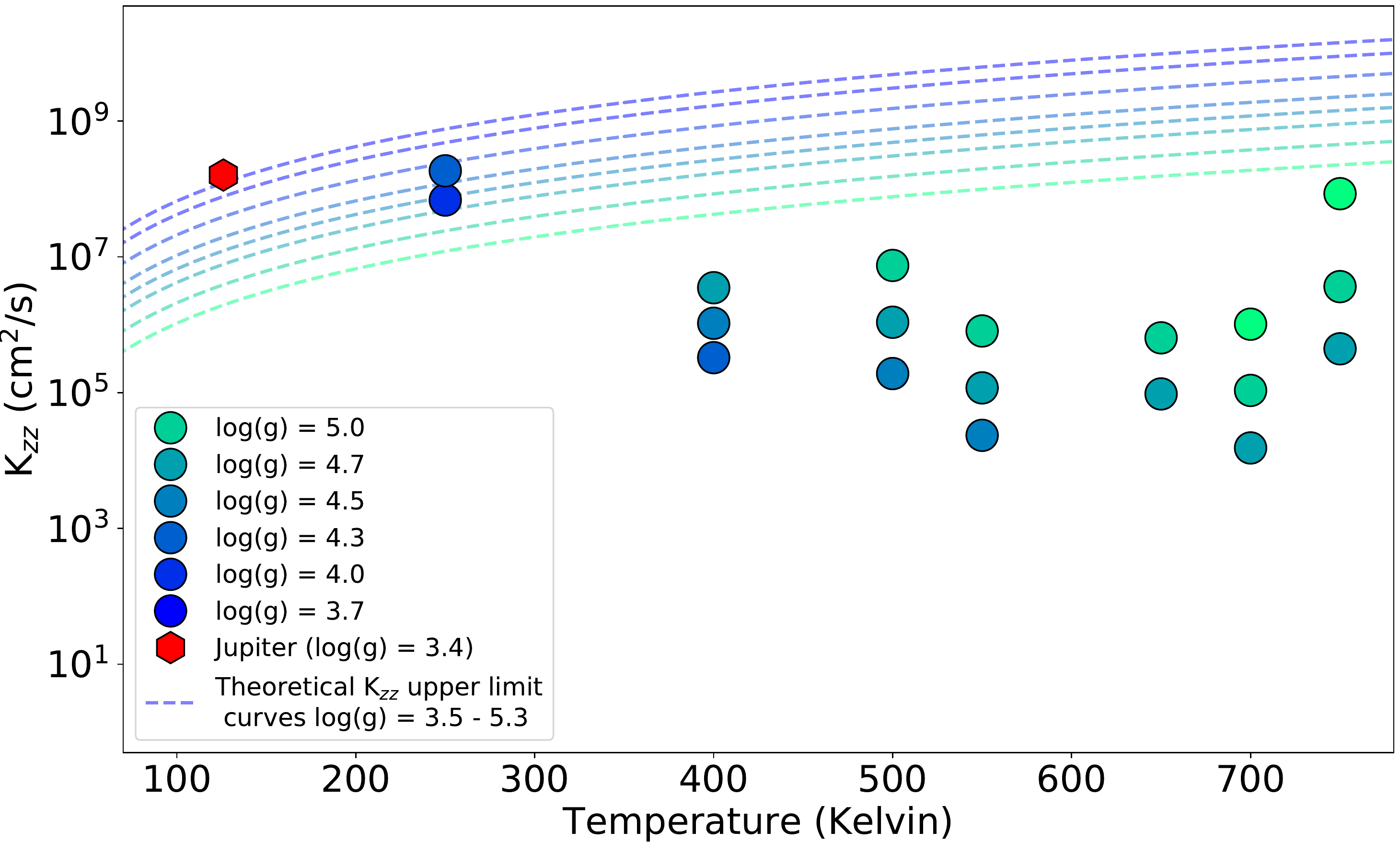}
\caption{The best fit $K_{zz}$ values versus temperature for the adopted models with different allowable surface gravities according to the Sonora Bobcat evolution models. UGPS 0722 (550 K) is plotted as if its allowable age range is 1 - 10 Gyr old. Hotter substellar objects have a wider range of possible surface gravities creating a larger spread in estimated possible $K_{zz}$ values. The dashed lines are the theoretical upper limits of $K_{zz}$ assuming all of the energy from internal heat drives convection. From top to bottom, the theoretical curves correspond to surface gravities of log(g) equal to 3.5, 3.7, 4.0, 4.3, 4.5, 4.7, 5.0, 5.3.}
\label{fig:kzz-temp}
\end{figure*}

\begin{figure*}
\centering
\includegraphics[width=6in]{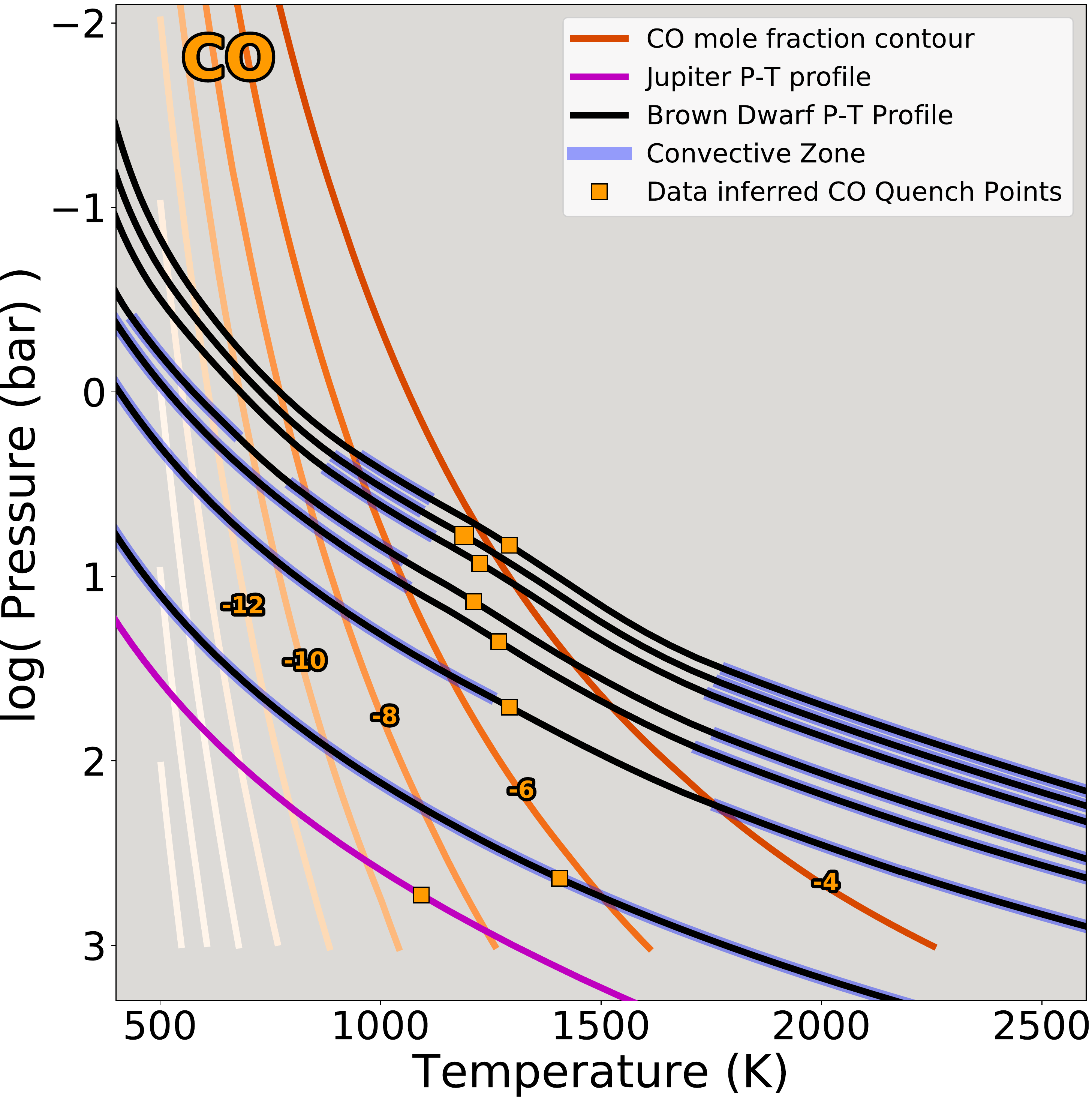}
\caption{Similar set up to Figure~\ref{fig:quench-co}, except the dots of constant K$_{zz}$ are removed and the model defined convective zones are highlighted in blue. For Jupiter and WISE 0855, the inferred quench points lie within convection dominated zones, but for the warmer objects the quench points can occur within a radiation dominated zone.}
\label{fig:convection-zones}
\end{figure*}

 The presence of detached convective zones within our brown dwarfs is explored by estimating the maximum value of K$_{zz}$ possible. We calculate  K$_{zz}$  using Equation 4  from \cite{2014ApJ...797...41Z}, which assumes an object's intrinsic flux the only driving force behind convection. A K$_{zz}$ maximum is estimated for a given effective temperature and surface gravity, then compared to the CO-inferred K$_{zz}$ values in Figure~\ref{fig:kzz-temp}. Jupiter and WISE 0855 have CO-inferred K$_{zz}$ values that are near the estimated maximum mixing rate. The warmer ($>$ 400 K) brown dwarfs have K$_{zz}$ values that are up to five factors of 10 below the estimated maximum mixing rate. The CO quench points of the warmer brown dwarfs lie within the radiative zones (Figure \ref{fig:convection-zones}), which could be suppressing the rate of vertical mixing below the theoretical maximum.  At 400 K and assuming a log(g) = 4.5, WISE 1541 is supposed to have two convective zones separated by a radiative layer, but uncertainties in gravity and temperature can make the difference between being fully convective and having detached convective zones split by radiative zones. Measured K$_{zz}$ values that are significantly below the theoretical upper limit could be a signature for detached convective zones if this trend is verified in more brown dwarfs and other directly imaged gas giants. 
 
 The brown dwarfs in our sample are assumed to resemble the atmospheres of widely separated gas giant exoplanets. Carbon monoxide absorption across the $M$-band can constrain the atmospheric quenching of gas giant exoplanets, but it also reduces their detectability as shown in Figure~\ref{fig:co-change}, where larger amounts of CO diminishes the total flux across the $M$-band (See Figure 4. \citealt{2003IAUS..211..345S}). Cold, directly imaged exoplanets and wide companions may have the advantage of being associated with a host star that can be aged and prospects for dynamical mass constraints. This information could reduce the parameter space over which K$_{zz}$ can be estimated for these worlds and maybe even provide the measurements needed to understand the fundamental driving forces behind atmospheric mixing.
 
\begin{deluxetable*}{llclcccc}
\caption{}
\tablecaption{Summary of Inferred and Measured Chemical Abundances}
\tablecomments{Summary of chemical abundances for our sample assuming the adopted temperatures and a log(g) = 4.5 for the brown dwarfs. All values the exponents of the log value. \ce{CO} abundances are directly measured from the spectra or previously published (a - \cite{2002Icar..159...95B}). The \ce{PH3} and \ce{NH3} abundances of Jupiter are compiled in this link \href{https://www.geochemsoc.org/publications/geochemicalnews/gn142jan10/atmosphericchemistryoftheg}{here}. The inferred \ce{NH3} abundances are from \cite{2019ApJ...877...24Z}. WISE 0855 has an estimate of \ce{PH3} not based on \ce{CO} from the paper \cite{2018ApJ...858...97M}.}
\tabletypesize{\footnotesize} 
\tablehead{
\colhead{Object} &  \colhead{Temp}   & \colhead{best fit CO}      & \colhead{Inferred K$_{zz}$}  & \colhead{Inferred \ce{PH3}} & \colhead{Published \ce{PH3}}  & \colhead{Inferred \ce{NH3}}  & \colhead{Published \ce{NH3}} \\
\colhead{}       &  \colhead{(K)}    & \colhead{log mole fraction}    & \colhead{cm$^{2}$ s$^{-1}$}  & \colhead{log mole fraction}     & \colhead{log mole fraction}      & \colhead{log mole fraction} & \colhead{log mole fraction}
}
\startdata
Gl 570 D    & 750 K  & -4.0  & 4.9 &  -6.72   & -           & -5.3   & -\\
2MASS J0415 & 700 K  & -4.5  & 3.6 &  -6.97   & -           & -5.2   & -     \\
WISE 0313   & 600 K  & -4.5  & 4.3 &  -6.80   & -           & -5.1   & -     \\
UGPS 0722   & 550 K  & -5.0  & 4.4 &  -6.73   & -           & -4.9   & -     \\
WISE 2056   & 500 K  & -5.0  & 5.3 &  -6.58   & -           & -4.8   & -4.44 \\
WISE 1541   & 400 K  & -5.5  & 6.0 &  -6.26   & -           & -4.5   & -4.43 \\
WISE 0855   & 250 K  & -6.5  & 8.5 & $<$-6.25 & $<$-6.30    & -3.9   &   -    \\
Jupiter     & 126 K  & -9.0$^{a}$&  8.2     &  -       & -5.96       &  -         & -3.2
\label{tbl:quench abundances}
\end{deluxetable*}

\subsection{Phosphine}
\label{sec:PH3}

\begin{figure*}
\centering
\includegraphics[width=6in]{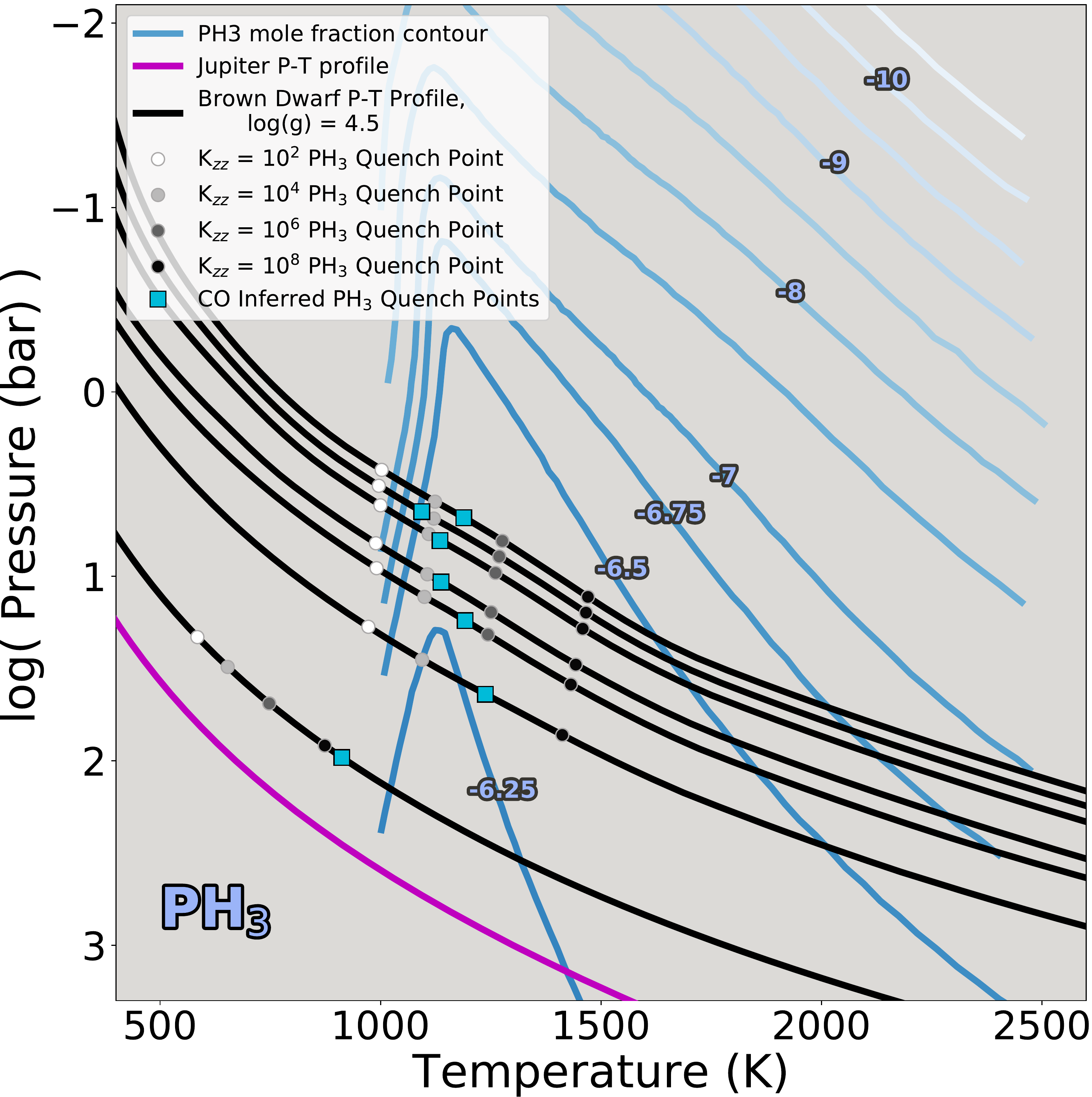}
\caption{Similar labeling convention as in Figure~\ref{fig:quench-co}, but the blue contour lines are the mole fractions of phosphine (\ce{PH3}) and the blue squares/triangle represent the inferred quench points of \ce{PH3} based on the $K_{zz}$ values calculated from CO. All of the brown dwarfs should have relatively high values of \ce{PH3} in their atmosphere as predicted in \cite{2014ApJ...787...78M}, yet the $L$ and $M$-band spectra do not show strong \ce{PH3} absorption. Jupiter's \ce{PH3} quench points occur in lower pressure and temperature regions outside of this plot}
\label{fig:quench-ph3}
\end{figure*}

Phosphine (\ce{PH3}) is a signature of disequilibrium chemistry that has been detected within Jupiter's $L$ and $M$-band spectra. Using the K$_{zz}$ values inferred from CO, we estimate the expected abundances of \ce{PH3} for the brown dwarfs and Jupiter. The locations of the \ce{PH3} quench points are shown in Figure ~\ref{fig:quench-ph3} and the expected \ce{PH3} values have mole fractions from 10$^{-6.7}$ to 10$^{-6.25}$, quenching at pressures similar to CO. The spread of possible \ce{PH3} abundances are not as large as CO, but the feature should be observable at mole fractions of 10$^{-7}$ in low resolution spectra from 4.5 $\mu$m to 4.65 $\mu$m ($M$-band) and 4.05 $\mu$m to 4.10 $\mu$m ($L$-band) \citep{2018ApJ...858...97M}. The vertical mixing inferred from CO tells us that \ce{PH3} should be observable in all of our brown dwarfs, but the wavelengths where \ce{PH3} could be observable occurs in the lowest transmission and lowest signal to noise regions of the spectra. Phosphine absorption is not obvious by eye, however more detailed analysis can place an upper limit on abundances. WISE 0855 should have abundances of \ce{PH3} that are detectable within the Gemini/GNIRS spectra according to estimates in \cite{2018ApJ...858...97M}, but it was not seen in the data. \cite{2017ApJ...850..199W} also concluded that a 500 K effective temperature brown dwarf should have a \ce{PH3} mole fraction of 10$^{-6.5}$ down to at least log(P) = 3, assuming a K$_{zz}$ of 10$^{9}$ cm$^{2}$ s$^{-1}$. Between K$_{zz}$ values of 10$^{2}$ to 10$^{8}$ the predicted disequilibrium abundance of \ce{PH3} can only change by a factor of 10 for brown dwarfs with effective temperatures of 400 K and above. The abundance contours of \ce{PH3} shown in Figure ~\ref{fig:quench-ph3} do cross some pressure-temperature profiles twice, which could lead to degenerate measurements of K$_{zz}$ if \ce{PH3} is the only molecule measured. This is not the case for \ce{CO} because carbon monoxide abundance increases with both temperature and pressure.

\subsection{Ammonia}
\label{sec:NH3}
\begin{figure*}
\centering
\includegraphics[width=6in]{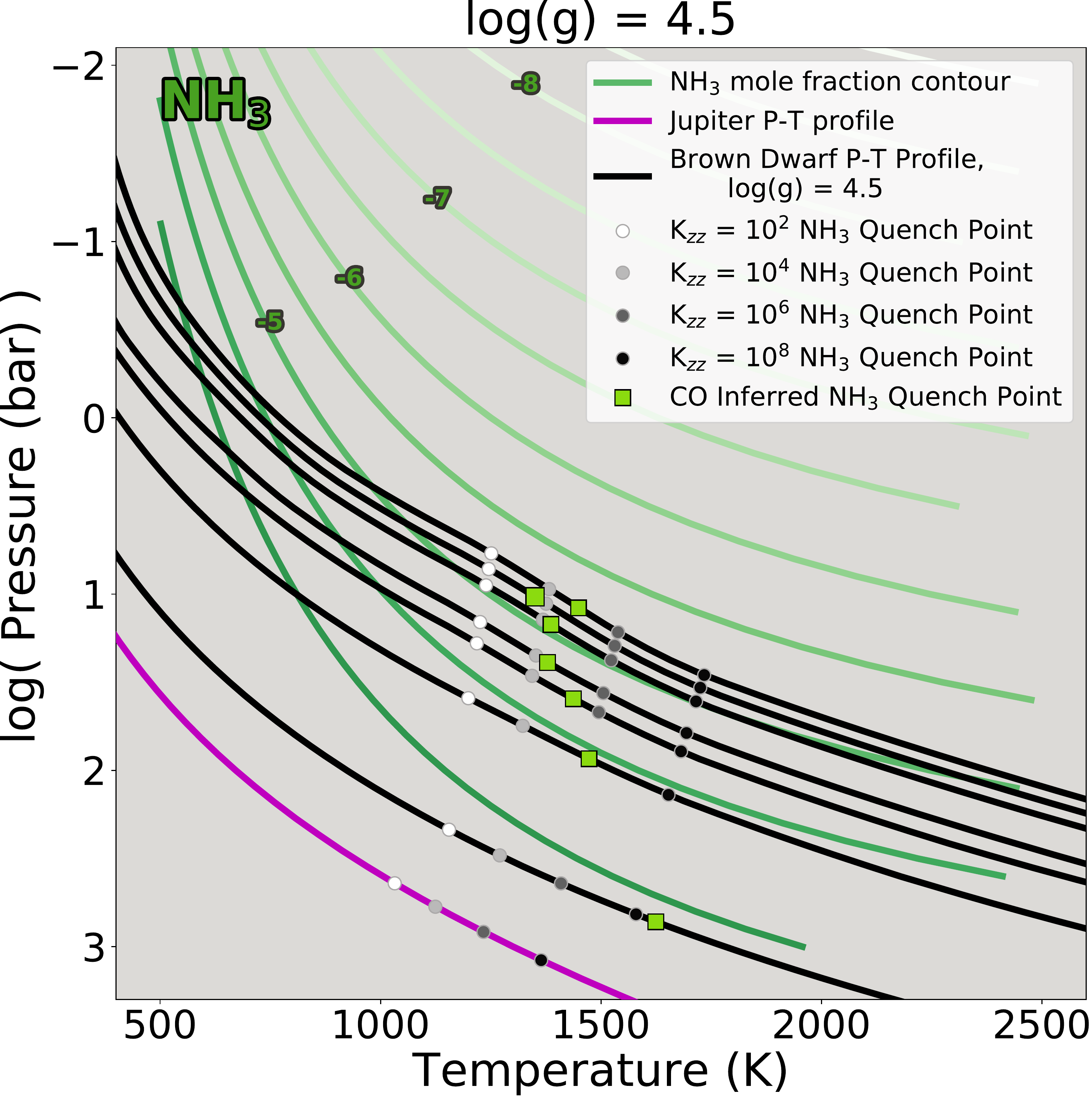}
\caption{Similar labeling convention as in Figure~\ref{fig:quench-co}, but the green contour lines are the mole fractions of ammonia (\ce{NH3}) and the green squares represent the inferred quench points of \ce{NH3} based on the $K_{zz}$ values calculated from CO. The \ce{NH3} mole fraction contours run almost parallel with some of brown dwarf pressure-temperature profiles, therefore \ce{NH3} abundances are somewhat insensitive to atmospheric mixing over the typical $K_{zz}$ values in a gas-giant object. }
\label{fig:quench-nh3}
\end{figure*}

The onset of ammonia (\ce{NH3}) is supposed to be a defining feature of Y-dwarfs because of their cooler effective temperatures and like \ce{PH3} it offers another element to understand the composition of brown dwarfs and exoplanets. Disequilibrium abundances of \ce{NH3} have been previously detected in near- and mid-infrared brown dwarfs spectra \citep{2007ApJ...656.1136S, 2009ApJ...695..844G, 2010ApJ...720..252L,2015ApJ...799...37L}. We also estimate the abundances of \ce{NH3} based on the K$_{zz}$ values inferred from \ce{CO}. The predicted quench points of \ce{NH3} for specific values of K$_{zz}$ from \ce{CO} are shown in Figure ~\ref{fig:quench-nh3}. Below a pressure of 1 bar, the contour lines of \ce{NH3} abundances are almost parallel with the pressure temperature profiles. This implies that \ce{NH3} is a strong indicator of effective temperature and comparatively less sensitive to atmospheric quenching with respect to \ce{CO} and \ce{PH3}, which is in agreement with the results of \cite{2006ApJ...647..552S}. From K$_{zz}$ values of 10$^{2}$ cm$^{2}$ s$^{-1}$  to 10$^{8}$ cm$^{2}$ s$^{-1}$ the abundance stays within a factor of 10 for warmer objects, but this may be less true for cooler objects such as Jupiter. The estimates for \ce{NH3} (Table ~\ref{tbl:quench abundances} based on \ce{CO} are consistent with the \ce{NH3} abundances derived from retrievals for WISE 2056 and WISE 1541 computed in \cite{2019ApJ...877...24Z}.

At equilibrium, the nitrogen abundance is primarily dictated by the balance between \ce{N2} and \ce{NH3}, but there are minor nitrogen bearing gases such as \ce{HCN} that affects where when the quenching point occurs at  high pressures and temperatures \citep{2002Icar..155..393L, 2014ApJ...797...41Z}. Minor nitrogen bearing gases are only expected to change the abundance of \ce{NH3} by $\sim$ 10\% or so, which may be difficult to detect with the ground based data that is typically published, but an effect to consider in the high quality spectra that will be taken with JWST. \ce{HCN} does have distinct opacity signatures across the 3 $\mu$m - 5 $\mu$m range as shown in \cite{2018ApJ...858...97M}, but the opacity per molecule is strongest at longer wavelengths with strong signatures at $\sim$7$\mu$m and $\sim$14$\mu$m. MIRI/JWST and longer wavelength instruments offer opportunities to understand the nitrogen chemical networks in brown dwarfs and gas giant exoplanets.

\section{Summary}
In this paper we combine new Gemini/GNIRS data and previously published $M$-band spectra to create a temperature sequence of 8 gaseous objects: Gl 570D, 2M 0415, UGPS 0722, WISE 0313, WISE 2056, WISE 1541, WISE 0855, and Jupiter. Our sample covers 750 K to 126 K, starting with brown dwarfs within the effective temperature range of the current coldest directly imaged companions (ex. 51 Eri b, GJ 504b), extending into the regime of could-be discovered exoplanets, and ending with our Solar System's largest gas giant, Jupiter. Enhanced abundances of carbon monoxide are detected in all of our objects and mole fractions between 10$^{-4}$ - 10$^{-7}$ are needed to better fit the spectra.  Clouds and enhanced carbon monoxide were shown to improve the spectral fitting for WISE 0855.

The best fit \ce{CO} values are evidence for convection driven, disequilibrium chemistry and the eddy diffusion coefficient (K$_{zz}$) is estimated for each object based on those abundances. The estimated K$_{zz}$ values of the sample spans from 10$^{4}$ cm$^{2}$ s$^{-1}$ to 10$^{8.5}$ cm$^{2}$ s$^{-1}$. The coolest objects in our sample, WISE 0855 and Jupiter, have estimated mixing strengths close to their theoretical upper limit, while brown dwarfs 400 K and warmer mix below this limit. This may be due to the presence of a predicted radiative zone at temperatures greater than 1100 K in the warmer atmosphere. Brown dwarfs and directly imaged gas giant planets are mostly convective, but warmer, higher gravity objects have detached convective zones, where mixing could be less efficient within the radiative layer and this is supported by our observations.

Using the estimated K$_{zz}$ values from CO, predictions are made for phosphine and ammonia. All of the brown dwarfs should have detectable mole fractions of \ce{PH3} between 10$^{-6.7}$ to 10$^{-6.25}$, but none of them show obvious absorbprtion within the $M$-band or previously published $L$-band data. In addition to this, we may not understand the behavior of phosphorous bearing molecules deep in the atmosphere of brown dwarfs. Ammonia is relatively insensitive to atmospheric mixing and our values of predicted \ce{NH3} based on \ce{CO} estimates are consistent with the retrieval analysis published in \cite{2019ApJ...877...24Z}. The chemical abundances of \ce{PH3} and \ce{NH3} will be best constrained with the medium resolution (R$\sim$1,000), high signal-to-noise spectroscopy that can be achieved with JWST.

Directly imaged, gas giant exoplanets are often targeted while young, have lower surface gravities, and condense clouds at lower pressures than similar effective temperature brown dwarfs. Because of this, directly imaged exoplanets may be susceptible to atmospheric quenching and a delayed onset temperature for water clouds compared to brown dwarfs \citep{2014ApJ...787...78M}. The prevalence of \ce{CO} in objects that are likely older than a 1 Gyr old suggests that disequilibrium \ce{CO} absorption may be common in colder gas giant exoplanets. Excess \ce{CO} absorption not only has implications for atmospheric characterization, but the overall detectability of these objects in planet finding surveys. More robust K$_{zz}$ estimates and tests of atmospheric physics could be made if larger numbers of substellar companions and directly imaged planets with age or dynamical mass constraints can be characterized. Cool brown dwarfs are excellent testing grounds for refining our atmospheric models and making predictions for observations of directly imaged exoplanets. Future mid-infrared spectroscopic studies of brown dwarfs and directly imaged gas giants will continue to expand our understanding of atmospheric chemistry and physics in gaseous objects.

\section{Acknowledgements}

The authors wish to recognize and acknowledge the very
significant cultural role and reverence that the summit of
Maunakea has always had within the indigenous Hawaiian
community. We are conducting observations from this
mountain, which is colonized land. 

This work benefited from
the Exoplanet Summer Program in the Other Worlds
Laboratory (OWL) at the University of California, Santa Cruz,
a program funded by the Heising-Simons Foundation. This research has also benefitted from the Y Dwarf Compendium maintained by Michael Cushing at https://sites.google.com/view/ydwarfcompendium/. We thank Davy Kirkpatrick and Sandy Leggett for providing unpublished parallaxes during the observation planning phase of this project. We thank Didier Saumon for feedback.

This work is based on observations obtained at the Gemini Observatory, which is operated by the Association of Universities for Research in Astronomy, Inc., under a cooperative agreement with the NSF on behalf of the Gemini partnership: the National Science Foundation (United States), the National Research Council (Canada), CONICYT (Chile), Ministerio de Ciencia, Tecnología e Innovación Productiva (Argentina), and Ministério da Ciência, Tecnologia e Inovação (Brazil). 

The observations presented in this work were difficult and often required real time changes, despite Gemini Observatory operating in queue. B.E.M. would like to thank the following Gemini Observatory staff and observers for their work and patience: J. Ball, K. Chiboucas, L. Fuhrman, M. Hoenig, S. Leggett, M. Lundquist, M. Marsset, J. Miller, A. Nitta, S. Pakzad, M. Pohlen, J. Scharwaechter, M. Schwamb, J. Shih, M. Simunovic, O. Smirnova, A. Smith, A. Stephens, S. Stewart. A.J.I.S. recognizes the NSF Planetary Astronomy, Award 1614320, which funded this work. B.E.M. is also thankful for the invaluable emotional support from family, friends, and colleagues.

\bibliographystyle{apj}


\begin{thebibliography}{}
\expandafter\ifx\csname natexlab\endcsname\relax\def\natexlab#1{#1}\fi

\bibitem[{{Ackerman} \& {Marley}(2001)}]{2001ApJ...556..872A}
{Ackerman}, A.~S., \& {Marley}, M.~S. 2001, \apj, 556, 872

\bibitem[{{Allers} \& {Liu}(2013)}]{2013ApJ...772...79A}
{Allers}, K.~N., \& {Liu}, M.~C. 2013, \apj, 772, 79

\bibitem[{{Barman} {et~al.}(2011){Barman}, {Macintosh}, {Konopacky}, \&
  {Marois}}]{2011ApJ...735L..39B}
{Barman}, T.~S., {Macintosh}, B., {Konopacky}, Q.~M., \& {Marois}, C. 2011,
  \apjl, 735, L39

\bibitem[{{Beichman} {et~al.}(2014){Beichman}, {Gelino}, {Kirkpatrick},
  {Cushing}, {Dodson-Robinson}, {Marley}, {Morley}, \&
  {Wright}}]{2014ApJ...783...68B}
{Beichman}, C., {Gelino}, C.~R., {Kirkpatrick}, J.~D., {et~al.} 2014, \apj,
  783, 68

\bibitem[{{B{\'e}zard} {et~al.}(2002){B{\'e}zard}, {Lellouch}, {Strobel},
  {Maillard}, \& {Drossart}}]{2002Icar..159...95B}
{B{\'e}zard}, B., {Lellouch}, E., {Strobel}, D., {Maillard}, J.-P., \&
  {Drossart}, P. 2002, \icarus, 159, 95

\bibitem[{{Bowler}(2016)}]{2016PASP..128j2001B}
{Bowler}, B.~P. 2016, \pasp, 128, 102001

\bibitem[{{Burgasser} {et~al.}(2006){Burgasser}, {Geballe}, {Leggett},
  {Kirkpatrick}, \& {Golimowski}}]{2006ApJ...637.1067B}
{Burgasser}, A.~J., {Geballe}, T.~R., {Leggett}, S.~K., {Kirkpatrick}, J.~D.,
  \& {Golimowski}, D.~A. 2006, \apj, 637, 1067

\bibitem[{{Burgasser} {et~al.}(2000){Burgasser}, {Kirkpatrick}, {Cutri},
  {McCallon}, {Kopan}, {Gizis}, {Liebert}, {Reid}, {Brown}, {Monet}, {Dahn},
  {Beichman}, \& {Skrutskie}}]{2000ApJ...531L..57B}
{Burgasser}, A.~J., {Kirkpatrick}, J.~D., {Cutri}, R.~M., {et~al.} 2000, \apjl,
  531, L57

\bibitem[{{Burrows} {et~al.}(2003){Burrows}, {Sudarsky}, \&
  {Lunine}}]{2003ApJ...596..587B}
{Burrows}, A., {Sudarsky}, D., \& {Lunine}, J.~I. 2003, \apj, 596, 587

\bibitem[{{Chabrier} {et~al.}(2019){Chabrier}, {Mazevet}, \&
  {Soubiran}}]{2019ApJ...872...51C}
{Chabrier}, G., {Mazevet}, S., \& {Soubiran}, F. 2019, \apj, 872, 51

\bibitem[{{Cushing} {et~al.}(2011){Cushing}, {Kirkpatrick}, {Gelino},
  {Griffith}, {Skrutskie}, {Mainzer}, {Marsh}, {Beichman}, {Burgasser},
  {Prato}, {Simcoe}, {Marley}, {Saumon}, {Freedman}, {Eisenhardt}, \&
  {Wright}}]{2011ApJ...743...50C}
{Cushing}, M.~C., {Kirkpatrick}, J.~D., {Gelino}, C.~R., {et~al.} 2011, \apj,
  743, 50

\bibitem[{{Cutri} \& {et al.}(2013)}]{2013yCat.2328....0C}
{Cutri}, R.~M., \& {et al.} 2013, VizieR Online Data Catalog, II/328

\bibitem[{{Dupuy} \& {Liu}(2012)}]{2012ApJS..201...19D}
{Dupuy}, T.~J., \& {Liu}, M.~C. 2012, \apjs, 201, 19

\bibitem[{{Elias} {et~al.}(2006){Elias}, {Joyce}, {Liang}, {Muller}, {Hileman},
  \& {George}}]{2006SPIE.6269E..4CE}
{Elias}, J.~H., {Joyce}, R.~R., {Liang}, M., {et~al.} 2006, in \procspie, Vol.
  6269, Society of Photo-Optical Instrumentation Engineers (SPIE) Conference
  Series, 62694C

\bibitem[{{Encrenaz} {et~al.}(1996){Encrenaz}, {de Graauw}, {Schaeidt},
  {Lellouch}, {Feuchtgruber}, {Beintema}, {Bezard}, {Drossart}, {Griffin},
  {Heras}, {Kessler}, {Leech}, {Morris}, {Roelfsema}, {Roos-Serote}, {Salama},
  {Vandenbussche}, {Valentijn}, {Davis}, \& {Naylor}}]{1996A&A...315L.397E}
{Encrenaz}, T., {de Graauw}, T., {Schaeidt}, S., {et~al.} 1996, \aap, 315, L397

\bibitem[{{Esplin} {et~al.}(2016){Esplin}, {Luhman}, {Cushing},
  {Hardegree-Ullman}, {Trucks}, {Burgasser}, \&
  {Schneider}}]{2016ApJ...832...58E}
{Esplin}, T.~L., {Luhman}, K.~L., {Cushing}, M.~C., {et~al.} 2016, \apj, 832,
  58

\bibitem[{{Faherty} {et~al.}(2014){Faherty}, {Tinney}, {Skemer}, \&
  {Monson}}]{2014ApJ...793L..16F}
{Faherty}, J.~K., {Tinney}, C.~G., {Skemer}, A., \& {Monson}, A.~J. 2014,
  \apjl, 793, L16

\bibitem[{{Faherty} {et~al.}(2016){Faherty}, {Riedel}, {Cruz}, {Gagne},
  {Filippazzo}, {Lambrides}, {Fica}, {Weinberger}, {Thorstensen}, {Tinney},
  {Baldassare}, {Lemonier}, \& {Rice}}]{2016ApJS..225...10F}
{Faherty}, J.~K., {Riedel}, A.~R., {Cruz}, K.~L., {et~al.} 2016, \apjs, 225, 10

\bibitem[{{Filippazzo} {et~al.}(2015){Filippazzo}, {Rice}, {Faherty}, {Cruz},
  {Van Gordon}, \& {Looper}}]{2015ApJ...810..158F}
{Filippazzo}, J.~C., {Rice}, E.~L., {Faherty}, J., {et~al.} 2015, \apj, 810,
  158

\bibitem[{{Geballe} {et~al.}(2009){Geballe}, {Saumon}, {Golimowski}, {Leggett},
  {Marley}, \& {Noll}}]{2009ApJ...695..844G}
{Geballe}, T.~R., {Saumon}, D., {Golimowski}, D.~A., {et~al.} 2009, \apj, 695,
  844

\bibitem[{{Geballe} {et~al.}(2001){Geballe}, {Saumon}, {Leggett}, {Knapp},
  {Marley}, \& {Lodders}}]{2001ApJ...556..373G}
{Geballe}, T.~R., {Saumon}, D., {Leggett}, S.~K., {et~al.} 2001, \apj, 556, 373

\bibitem[{{Golimowski} {et~al.}(2004){Golimowski}, {Leggett}, {Marley}, {Fan},
  {Geballe}, {Knapp}, {Vrba}, {Henden}, {Luginbuhl}, {Guetter}, {Munn},
  {Canzian}, {Zheng}, {Tsvetanov}, {Chiu}, {Glazebrook}, {Hoversten},
  {Schneider}, \& {Brinkmann}}]{2004AJ....127.3516G}
{Golimowski}, D.~A., {Leggett}, S.~K., {Marley}, M.~S., {et~al.} 2004, \aj,
  127, 3516

\bibitem[{{Kim} {et~al.}(2015){Kim}, {Prato}, \&
  {McLean}}]{2015ascl.soft07017K}
{Kim}, S., {Prato}, L., \& {McLean}, I. 2015, {REDSPEC: NIRSPEC data
  reduction}, ascl:1507.017

\bibitem[{{Kippenhahn} {et~al.}(2012){Kippenhahn}, {Weigert}, \&
  {Weiss}}]{2012sse..book.....K}
{Kippenhahn}, R., {Weigert}, A., \& {Weiss}, A. 2012, {Stellar Structure and
  Evolution}, doi:10.1007/978-3-642-30304-3

\bibitem[{{Kirkpatrick}(2005)}]{2005ARA&A..43..195K}
{Kirkpatrick}, J.~D. 2005, \araa, 43, 195

\bibitem[{{Kirkpatrick} {et~al.}(2011){Kirkpatrick}, {Cushing}, {Gelino},
  {Griffith}, {Skrutskie}, {Marsh}, {Wright}, {Mainzer}, {Eisenhardt},
  {McLean}, {Thompson}, {Bauer}, {Benford}, {Bridge}, {Lake}, {Petty},
  {Stanford}, {Tsai}, {Bailey}, {Beichman}, {Bloom}, {Bochanski}, {Burgasser},
  {Capak}, {Cruz}, {Hinz}, {Kartaltepe}, {Knox}, {Manohar}, {Masters},
  {Morales-Calder{\'o}n}, {Prato}, {Rodigas}, {Salvato}, {Schurr}, {Scoville},
  {Simcoe}, {Stapelfeldt}, {Stern}, {Stock}, \& {Vacca}}]{2011ApJS..197...19K}
{Kirkpatrick}, J.~D., {Cushing}, M.~C., {Gelino}, C.~R., {et~al.} 2011, The
  Astrophysical Journal Supplement Series, 197, 19

\bibitem[{{Kirkpatrick} {et~al.}(2012){Kirkpatrick}, {Gelino}, {Cushing},
  {Mace}, {Griffith}, {Skrutskie}, {Marsh}, {Wright}, {Eisenhardt}, {McLean},
  {Mainzer}, {Burgasser}, {Tinney}, {Parker}, \&
  {Salter}}]{2012ApJ...753..156K}
{Kirkpatrick}, J.~D., {Gelino}, C.~R., {Cushing}, M.~C., {et~al.} 2012, \apj,
  753, 156

\bibitem[{{Kirkpatrick} {et~al.}(2019){Kirkpatrick}, {Martin}, {Smart},
  {Cayago}, {Beichman}, {Marocco}, {Gelino}, {Faherty}, {Cushing}, {Schneider},
  {Mace}, {Tinney}, {Wright}, {Lowrance}, {Ingalls}, {Vrba}, {Munn}, {Dahm}, \&
  {McLean}}]{2019ApJS..240...19K}
{Kirkpatrick}, J.~D., {Martin}, E.~C., {Smart}, R.~L., {et~al.} 2019, \apjs,
  240, 19

\bibitem[{{Knapp} {et~al.}(2004){Knapp}, {Leggett}, {Fan}, {Marley}, {Geballe},
  {Golimowski}, {Finkbeiner}, {Gunn}, {Hennawi}, {Ivezi{\'c}}, {Lupton},
  {Schlegel}, {Strauss}, {Tsvetanov}, {Chiu}, {Hoversten}, {Glazebrook},
  {Zheng}, {Hendrickson}, {Williams}, {Uomoto}, {Vrba}, {Henden}, {Luginbuhl},
  {Guetter}, {Munn}, {Canzian}, {Schneider}, \&
  {Brinkmann}}]{2004AJ....127.3553K}
{Knapp}, G.~R., {Leggett}, S.~K., {Fan}, X., {et~al.} 2004, \aj, 127, 3553

\bibitem[{{Leggett} {et~al.}(2015){Leggett}, {Morley}, {Marley}, \&
  {Saumon}}]{2015ApJ...799...37L}
{Leggett}, S.~K., {Morley}, C.~V., {Marley}, M.~S., \& {Saumon}, D. 2015, \apj,
  799, 37

\bibitem[{{Leggett} {et~al.}(2013){Leggett}, {Morley}, {Marley}, {Saumon},
  {Fortney}, \& {Visscher}}]{2013ApJ...763..130L}
{Leggett}, S.~K., {Morley}, C.~V., {Marley}, M.~S., {et~al.} 2013, \apj, 763,
  130

\bibitem[{{Leggett} {et~al.}(2010){Leggett}, {Saumon}, {Burningham}, {Cushing},
  {Marley}, \& {Pinfield}}]{2010ApJ...720..252L}
{Leggett}, S.~K., {Saumon}, D., {Burningham}, B., {et~al.} 2010, \apj, 720, 252

\bibitem[{{Leggett} {et~al.}(2017){Leggett}, {Tremblin}, {Esplin}, {Luhman}, \&
  {Morley}}]{2017ApJ...842..118L}
{Leggett}, S.~K., {Tremblin}, P., {Esplin}, T.~L., {Luhman}, K.~L., \&
  {Morley}, C.~V. 2017, \apj, 842, 118

\bibitem[{{Leggett} {et~al.}(2012){Leggett}, {Saumon}, {Marley}, {Lodders},
  {Canty}, {Lucas}, {Smart}, {Tinney}, {Homeier}, {Allard}, {Burningham},
  {Day-Jones}, {Fegley}, {Ishii}, {Jones}, {Marocco}, {Pinfield}, \&
  {Tamura}}]{2012ApJ...748...74L}
{Leggett}, S.~K., {Saumon}, D., {Marley}, M.~S., {et~al.} 2012, \apj, 748, 74

\bibitem[{{Leggett} {et~al.}(2019{\natexlab{a}}){Leggett}, {Dupuy}, {Morley},
  {Marley}, {Best}, {Liu}, {Apai}, {Casewell}, {Geballe}, {Gizis}, {Pineda},
  {Rieke}, \& {Wright}}]{2019ApJ...882..117L}
{Leggett}, S.~K., {Dupuy}, T.~J., {Morley}, C.~V., {et~al.} 2019{\natexlab{a}},
  \apj, 882, 117

\bibitem[{{Leggett} {et~al.}(2019{\natexlab{b}}){Leggett}, {Dupuy}, {Morley},
  {Marley}, {Best}, {Liu}, {Apai}, {Casewell}, {Geballe}, {Gizis}, {Pineda},
  {Rieke}, \& {Wright}}]{2019arXiv190707798L}
---. 2019{\natexlab{b}}, arXiv e-prints, arXiv:1907.07798

\bibitem[{{Li} {et~al.}(2012){Li}, {Baines}, {Smith}, {West},
  {P{\'e}rez-Hoyos}, {Trammell}, {Simon-Miller}, {Conrath}, {Gierasch},
  {Orton}, {Nixon}, {Filacchione}, {Fry}, \& {Momary}}]{2012JGRE..11711002L}
{Li}, L., {Baines}, K.~H., {Smith}, M.~A., {et~al.} 2012, Journal of
  Geophysical Research (Planets), 117, E11002

\bibitem[{{Line} {et~al.}(2015){Line}, {Teske}, {Burningham}, {Fortney}, \&
  {Marley}}]{2015ApJ...807..183L}
{Line}, M.~R., {Teske}, J., {Burningham}, B., {Fortney}, J.~J., \& {Marley},
  M.~S. 2015, \apj, 807, 183

\bibitem[{{Line} {et~al.}(2017){Line}, {Marley}, {Liu}, {Burningham}, {Morley},
  {Hinkel}, {Teske}, {Fortney}, {Freedman}, \& {Lupu}}]{2017ApJ...848...83L}
{Line}, M.~R., {Marley}, M.~S., {Liu}, M.~C., {et~al.} 2017, \apj, 848, 83

\bibitem[{{Lodders} \& {Fegley}(2002)}]{2002Icar..155..393L}
{Lodders}, K., \& {Fegley}, B. 2002, \icarus, 155, 393

\bibitem[{{Lucas} {et~al.}(2010){Lucas}, {Tinney}, {Burningham}, {Leggett},
  {Pinfield}, {Smart}, {Jones}, {Marocco}, {Barber}, {Yurchenko}, {Tennyson},
  {Ishii}, {Tamura}, {Day-Jones}, {Adamson}, {Allard}, \&
  {Homeier}}]{2010MNRAS.408L..56L}
{Lucas}, P.~W., {Tinney}, C.~G., {Burningham}, B., {et~al.} 2010, \mnras, 408,
  L56

\bibitem[{{Luhman}(2014)}]{2014ApJ...786L..18L}
{Luhman}, K.~L. 2014, \apjl, 786, L18

\bibitem[{{Luhman} \& {Esplin}(2016)}]{2016AJ....152...78L}
{Luhman}, K.~L., \& {Esplin}, T.~L. 2016, \aj, 152, 78

\bibitem[{{Martin} {et~al.}(2017){Martin}, {Mace}, {McLean}, {Logsdon}, {Rice},
  {Kirkpatrick}, {Burgasser}, {McGovern}, \& {Prato}}]{2017ApJ...838...73M}
{Martin}, E.~C., {Mace}, G.~N., {McLean}, I.~S., {et~al.} 2017, \apj, 838, 73

\bibitem[{{Miles} {et~al.}(2018){Miles}, {Skemer}, {Barman}, {Allers}, \&
  {Stone}}]{2018ApJ...869...18M}
{Miles}, B.~E., {Skemer}, A.~J., {Barman}, T.~S., {Allers}, K.~N., \& {Stone},
  J.~M. 2018, \apj, 869, 18

\bibitem[{{Morley} {et~al.}(2015){Morley}, {Fortney}, {Marley}, {Zahnle},
  {Line}, {Kempton}, {Lewis}, \& {Cahoy}}]{2015ApJ...815..110M}
{Morley}, C.~V., {Fortney}, J.~J., {Marley}, M.~S., {et~al.} 2015, \apj, 815,
  110

\bibitem[{{Morley} {et~al.}(2014){Morley}, {Marley}, {Fortney}, {Lupu},
  {Saumon}, {Greene}, \& {Lodders}}]{2014ApJ...787...78M}
{Morley}, C.~V., {Marley}, M.~S., {Fortney}, J.~J., {et~al.} 2014, \apj, 787,
  78

\bibitem[{{Morley} {et~al.}(2018){Morley}, {Skemer}, {Allers}, {Marley},
  {Faherty}, {Visscher}, {Beiler}, {Miles}, {Lupu}, {Freedman}, {Fortney},
  {Geballe}, \& {Bjoraker}}]{2018ApJ...858...97M}
{Morley}, C.~V., {Skemer}, A.~J., {Allers}, K.~N., {et~al.} 2018, \apj, 858, 97

\bibitem[{{Noll}(1993)}]{1993ASPC...41...29N}
{Noll}, K.~S. 1993, in Astronomical Society of the Pacific Conference Series,
  Vol.~41, Astronomical Infrared Spectroscopy: Future Observational Directions,
  ed. S.~{Kwok}, 29

\bibitem[{{Noll} {et~al.}(1997){Noll}, {Geballe}, \&
  {Marley}}]{1997ApJ...489L..87N}
{Noll}, K.~S., {Geballe}, T.~R., \& {Marley}, M.~S. 1997, \apjl, 489, L87

\bibitem[{{Oppenheimer} {et~al.}(1998){Oppenheimer}, {Kulkarni}, {Matthews}, \&
  {van Kerkwijk}}]{1998ApJ...502..932O}
{Oppenheimer}, B.~R., {Kulkarni}, S.~R., {Matthews}, K., \& {van Kerkwijk},
  M.~H. 1998, \apj, 502, 932

\bibitem[{{Patten} {et~al.}(2006){Patten}, {Stauffer}, {Burrows}, {Marengo},
  {Hora}, {Luhman}, {Sonnett}, {Henry}, {Raghavan}, {Megeath}, {Liebert}, \&
  {Fazio}}]{2006ApJ...651..502P}
{Patten}, B.~M., {Stauffer}, J.~R., {Burrows}, A., {et~al.} 2006, \apj, 651,
  502

\bibitem[{{Saumon} {et~al.}(2006){Saumon}, {Marley}, {Cushing}, {Leggett},
  {Roellig}, {Lodders}, \& {Freedman}}]{2006ApJ...647..552S}
{Saumon}, D., {Marley}, M.~S., {Cushing}, M.~C., {et~al.} 2006, \apj, 647, 552

\bibitem[{{Saumon} {et~al.}(2003){Saumon}, {Marley}, {Lodders}, \&
  {Freedman}}]{2003IAUS..211..345S}
{Saumon}, D., {Marley}, M.~S., {Lodders}, K., \& {Freedman}, R.~S. 2003, in IAU
  Symposium, Vol. 211, Brown Dwarfs, ed. E.~{Mart{\'\i}n}, 345

\bibitem[{{Saumon} {et~al.}(2007){Saumon}, {Marley}, {Leggett}, {Geballe},
  {Stephens}, {Golimowski}, {Cushing}, {Fan}, {Rayner}, {Lodders}, \&
  {Freedman}}]{2007ApJ...656.1136S}
{Saumon}, D., {Marley}, M.~S., {Leggett}, S.~K., {et~al.} 2007, \apj, 656, 1136

\bibitem[{{Schneider} {et~al.}(2016){Schneider}, {Cushing}, {Kirkpatrick}, \&
  {Gelino}}]{2016ApJ...823L..35S}
{Schneider}, A.~C., {Cushing}, M.~C., {Kirkpatrick}, J.~D., \& {Gelino}, C.~R.
  2016, \apjl, 823, L35

\bibitem[{{Schneider} {et~al.}(2015){Schneider}, {Cushing}, {Kirkpatrick},
  {Gelino}, {Mace}, {Wright}, {Eisenhardt}, {Skrutskie}, {Griffith}, \&
  {Marsh}}]{2015ApJ...804...92S}
{Schneider}, A.~C., {Cushing}, M.~C., {Kirkpatrick}, J.~D., {et~al.} 2015,
  \apj, 804, 92

\bibitem[{{Skemer} {et~al.}(2016){Skemer}, {Morley}, {Allers}, {Geballe},
  {Marley}, {Fortney}, {Faherty}, {Bjoraker}, \& {Lupu}}]{2016ApJ...826L..17S}
{Skemer}, A.~J., {Morley}, C.~V., {Allers}, K.~N., {et~al.} 2016, \apj, 826,
  L17

\bibitem[{{Skrutskie} {et~al.}(2006){Skrutskie}, {Cutri}, {Stiening},
  {Weinberg}, {Schneider}, {Carpenter}, {Beichman}, {Capps}, {Chester},
  {Elias}, {Huchra}, {Liebert}, {Lonsdale}, {Monet}, {Price}, {Seitzer},
  {Jarrett}, {Kirkpatrick}, {Gizis}, {Howard}, {Evans}, {Fowler}, {Fullmer},
  {Hurt}, {Light}, {Kopan}, {Marsh}, {McCallon}, {Tam}, {Van Dyk}, \&
  {Wheelock}}]{2006AJ....131.1163S}
{Skrutskie}, M.~F., {Cutri}, R.~M., {Stiening}, R., {et~al.} 2006, \aj, 131,
  1163

\bibitem[{{Sorahana} \& {Yamamura}(2012)}]{2012ApJ...760..151S}
{Sorahana}, S., \& {Yamamura}, I. 2012, \apj, 760, 151

\bibitem[{{Thorngren} {et~al.}(2016){Thorngren}, {Fortney}, {Murray-Clay}, \&
  {Lopez}}]{2016ApJ...831...64T}
{Thorngren}, D.~P., {Fortney}, J.~J., {Murray-Clay}, R.~A., \& {Lopez}, E.~D.
  2016, \apj, 831, 64

\bibitem[{{Visscher} {et~al.}(2006){Visscher}, {Lodders}, \&
  {Fegley}}]{2006ApJ...648.1181V}
{Visscher}, C., {Lodders}, K., \& {Fegley}, Bruce, J. 2006, \apj, 648, 1181

\bibitem[{{Visscher} \& {Moses}(2011)}]{2011ApJ...738...72V}
{Visscher}, C., \& {Moses}, J.~I. 2011, \apj, 738, 72

\bibitem[{{Visscher} {et~al.}(2010){Visscher}, {Moses}, \&
  {Saslow}}]{2010Icar..209..602V}
{Visscher}, C., {Moses}, J.~I., \& {Saslow}, S.~A. 2010, \icarus, 209, 602

\bibitem[{{Wang} {et~al.}(2017){Wang}, {Miguel}, \&
  {Lunine}}]{2017ApJ...850..199W}
{Wang}, D., {Miguel}, Y., \& {Lunine}, J. 2017, \apj, 850, 199

\bibitem[{{Wright} {et~al.}(2014){Wright}, {Mainzer}, {Kirkpatrick}, {Masci},
  {Cushing}, {Bauer}, {Fajardo-Acosta}, {Gelino}, {Beichman}, {Skrutskie},
  {Grav}, {Eisenhardt}, \& {Cutri}}]{2014AJ....148...82W}
{Wright}, E.~L., {Mainzer}, A., {Kirkpatrick}, J.~D., {et~al.} 2014, \aj, 148,
  82

\bibitem[{{Zahnle} \& {Marley}(2014)}]{2014ApJ...797...41Z}
{Zahnle}, K.~J., \& {Marley}, M.~S. 2014, \apj, 797, 41

\bibitem[{{Zalesky} {et~al.}(2019){Zalesky}, {Line}, {Schneider}, \&
  {Patience}}]{2019ApJ...877...24Z}
{Zalesky}, J.~A., {Line}, M.~R., {Schneider}, A.~C., \& {Patience}, J. 2019,
  \apj, 877, 24

\end{thebibliography}
\end{document}